\documentclass[journal]{IEEEtran}

\usepackage{cite}
\usepackage{amsmath,amssymb,amsfonts}
\usepackage{algorithmic}
\usepackage{graphicx}
\usepackage{mathtools}
\usepackage{color,soul}
\usepackage{xcolor}
\usepackage{epstopdf}
\usepackage[tight]{subfigure}
\usepackage{acronym}

\acrodef{RIS}{Reconfigurable Intelligent Surface}
\acrodef{LIS}{Large Intelligent Surface}
\acrodef{SDM}{Software-Defined Metamaterial}
\acrodef{EM}{electromagnetic}
\acrodef{HSF}{HyperSurface}

\renewcommand{\hl}[1]{#1}

\begin{document}

	\title{Error Analysis of Programmable Metasurfaces\\for Beam Steering}

	\author{Hamidreza Taghvaee, Albert Cabellos-Aparicio, Julius Georgiou, and Sergi Abadal
		\thanks{H. Taghvaee, A. Cabellos-Aparicio, and S. Abadal are with the NaNoNetworking Center in Catalonia (N3Cat), Universitat Polit\`{e}cnica de Catalunya, 08034 Barcelona, Spain (e-mail: taghvaee@ac.upc.edu, acabello@ac.upc.edu, abadal@ac.upc.edu)}%
		\thanks{J. Georgiou is with the Department of Electrical and Computer Engineering, University of Cyprus, P.O. Box 20537, 1678, Nicosia, Cyprus (e-mail: julio@ucy.ac.cy)}
	}

	\maketitle

\begin{abstract}
Recent years have seen the emergence of programmable metasurfaces, where the user can modify the EM response of the device via software. Adding reconfigurability to the already powerful EM capabilities of metasurfaces opens the door to novel cyber-physical systems with exciting applications in domains such as holography, cloaking, or wireless communications. This paradigm shift, however, comes with a non-trivial increase of the complexity of the metasurfaces that will pose new reliability challenges stemming from the need to integrate tuning, control, and communication resources to implement the programmability. While metasurfaces will become prone to failures, little is known about their tolerance to errors. To bridge this gap, this paper examines the reliability problem in programmable metamaterials by proposing an error model and a general methodology for error analysis. To derive the error model, the causes and potential impact of faults are identified and discussed qualitatively. The methodology is presented and exemplified for beam steering, which constitutes a relevant case for programmable metasurfaces. Results show that performance degradation depends on the type of error and its spatial distribution and that, in beam steering, error rates over 20\% can still be considered acceptable.
\end{abstract}

\acresetall


\section{Introduction}
\label{sec:intro}

Metamaterials have garnered significant attention in the last decade as they enable unprecedented levels of \ac{EM} control \cite{Engheta2006} and have opened the door to disruptive advances across domains such as imaging, integrated optics, or wireless communications \cite{Glybovski2016, Taghvaee2014, Chen2016, Vellucci2017}. Metasurfaces, the thin-film analog of metamaterials, are generally comprised of a planar array of subwavelength elements over a substrate, i.e. the unit cells, and inherit the unique properties of their 3D counterparts while minimizing bulkiness, losses, and cost. Functionalities such as beam steering, focusing, vorticity control, or RCS reduction have been demonstrated across the spectrum, from microwaves \cite{Huang2017, Li2017b, Tcvetkova2018} to terahertz \cite{Qu2015, Hosseininejad2019, Liu2016a, Taghvaee2017, Qu2017}, or optical frequencies \cite{ChenXZ2012, Li2015}.

Early works in the field of metamaterials had two main drawbacks, namely, non-adaptivity and non-reconfigurability. This is because, due to their highly resonant nature, unit cells are generally designed for a particular \ac{EM} function and scope. To address this issue, tunable elements or materials have been introduced in the unit cell design loop in order to provide global or local reconfigurability \cite{Oliveri2015}. Further, recent years have seen the emergence of programmable metasurfaces, this is, metasurfaces that incorporate local tunability and digital logic to easily reconfigure the \ac{EM} behavior from the outside.

Two main approaches have been proposed for the implementation of programmable metasurfaces, namely, (i) by interfacing the tunable elements through an external Field-Programmable Gate Array (FPGA) \cite{Cui2014, Wan2016}, or (ii) by integrating sensors, control units, and actuators within the metasurface structure \cite{AbadalACCESS, Tasolamprou8788546, Liaskos2018a, PhysRevApplied.11.044024}.

\begin{figure}[!t]
		\vspace{-0.7cm}
	\includegraphics[width=\columnwidth]{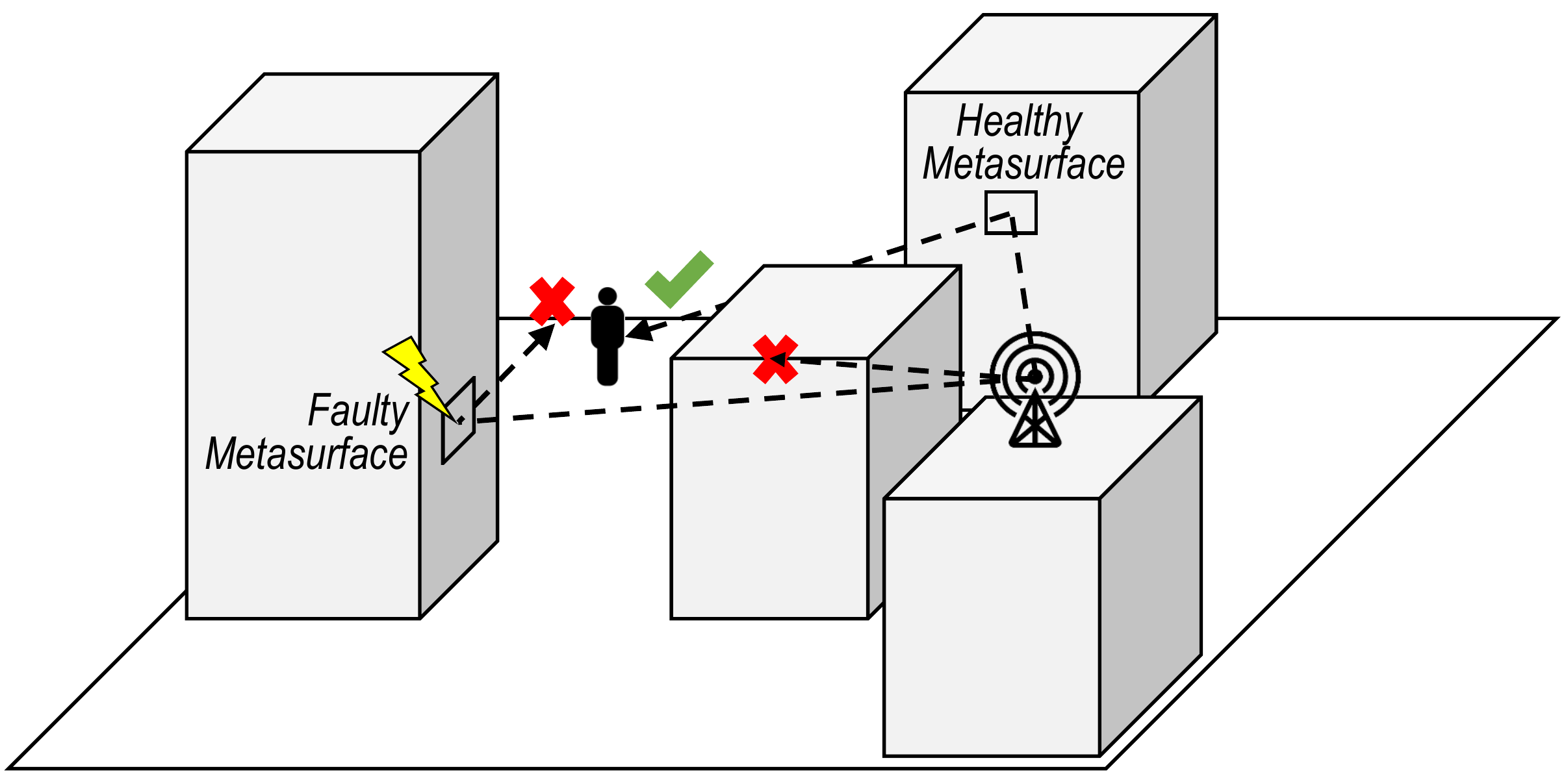} 

	\caption{Graphical illustration of an environment augmented with potentially faulty programmable metasurfaces. Since Line-of-Sight (LoS) propagation is not possible, the metasurfaces attempt to maximize the non-LoS power at the receiver by directing the reflections and altering the phase for coherent detection. A faulty metasurface may lead to service disruption by not pointing the reflected beam accurately.}
	\vspace{-0.4cm}
	\label{fig:intro}
\end{figure}

Programmable metasurfaces have opened the door to disruptive paradigms such as \acp{SDM} and \acp{RIS}, leading to the interconnection of metasurfaces, the use of machine learning and, eventually, the implementation of software-driven distributed intelligence on \ac{EM} control \cite{AbadalACCESS, Liaskos2015, di2019smart, huang2019reconfigurable}. This has potential to exert a disruptive impact in a plethora of application domains, including but not limited to holographic displays \cite{Li2017b}, stealth technology \cite{ma2019smart} or wireless communications \cite{Liaskos2018a, HsfNetworkTNET.2019}. In the latter case, the metasurfaces allow modifying the wireless channel by controlling the direction and phase of reflection. Figure \ref{fig:intro} illustrates a plausible scenario where service would normally be disrupted due to blockage, but that metasurfaces are capable of redress by directing the reflections to the user. This is a true paradigm shift for wireless communications, as the recent explosion of works can attest \cite{Liaskos2018a, Liaskos2018b, HsfNetworkTNET.2019, huang2019reconfigurable, perovic2019channel, bjornson2019intelligent, Basar2019}, because the wireless channel has traditionally been an inescapable limiting factor.

Alas, the transition from static to intelligent programmable metasurfaces has come at the cost of added design, fabrication, and embedding complexity. Programmable metasurfaces need to integrate tuning and control elements on a per-cell basis, electronic circuits to implement intelligence within the device, as well as mechanisms to interface the surface with the world. This poses important challenges at fabrication, calibration, deployment, and run time that, among others, affect reliability. In other words, metasurfaces will become prone to failure as they continue integrating sophisticated tuning, control and sensing circuits. However, the impact of faults on the performance of individual metasurfaces is not well understood yet.

We claim that error analysis is a necessary first step to understand the impact of transient or permanent failures on the performance of both a single metasurface and, crucially, a complete system. In the scenario of Figure \ref{fig:intro}, for instance, faults might lead to inaccuracies in the steering of the reflection and cause a drop in quality of service. Further, an error analysis would also allow to derive guidelines for the implementation of robust programmable metasurfaces, estimate the lifetime of the deployed ones, or even develop methods to save energy by power-gating a portion of the internal circuitry of the metasurface.

However, error analyses have not been carried out taking the particularities of metasurfaces into consideration. In \cite{badiu2019communication}, the authors evaluate the impact of phase errors in \ac{RIS} panels that shift the phase of impinging waves aiming for a coherent combination at the receiver. In that case, each \ac{RIS} is spaced apart and treated independently, ignoring the directions of impinging or reflected waves and limiting errors to system inaccuracies, i.e. quantization and estimation error, but no faults. This resembles the classical works that analyze the impact of errors in phased antenna arrays \cite{rondinelli1966effects, zahm1972effects, wang1992performance, poli2015dealing}. Such an analysis is therefore not directly applicable to metasurfaces, where (1) the causes of failures can be more varied due to the amount of control circuitry, and (2) the focus is on the impact of a number of failures (or the chained effect of a few of them) rather than on individual faulty components.

This paper proposes a framework to evaluate the impact of failures in programmable metasurfaces, distinguishing between the type of faults and their spatial distribution. Despite being applicable to any \ac{EM} functionality, here we use the methodology to study a beam steering metasurface at 26 GHz as a particular yet very relevant use case in metasurface-enabled 5G communications as shown in Fig. \ref{fig:intro} \cite{Liaskos2018a}. This contribution extends our previous work \cite{8702080} where the error model and the methodology were outlined, and a generic beam steering device was evaluated in a limited amount of cases. Here, we deepen the analysis by (i) introducing a realistic unit cell to improve the system model, (ii) exemplifying the effects of failures in the components of a tunable unit cell, (iii) evaluating the impact of faults in multiple performance metrics such as deviation from the target direction, and (iv) evaluating the impact of a much wider range of error type combinations.

The remainder of this paper is as follows. Section \ref{sec:background} provides background on programmable metasurfaces along with design factors such as coding and architecture. Section \ref{sec:Sys} introduces the metasurface model for the far field. Section \ref{sec:error} analyzes possible causes of errors and derives an error model. Section \ref{sec:methodology} presents the evaluation methodology, instantiated for the beam steering case. Section \ref{sec:results} shows the results of the analysis and Section \ref{sec:conclusions} concludes the paper.

\section{Background: Programmable Metasurfaces}
\label{sec:background}

Metasurfaces are structures generally composed of an array of subwavelength resonators referred to as \emph{unit cells}. The characteristics of these building blocks determine the metasurface response, this is, its absorption, reflection, and transmission characteristic \cite{Li2018, Chen2016, Tsilipakos2018a}. 

With the introduction of tunable or switchable elements, metasurfaces can adapt to different environments or take multiple functionalities \cite{Oliveri2015, Zhang2017, Tsilipakos2018, Taghvaee2017}. Tuning has been demonstrated in several forms, including thermal \cite{Lewi2019}, electrical \cite{Ju2011, Hosseininejad2019}, and optical tuning \cite{Zhao2015}; or the use pin diodes \cite{Yang2016}, varactors \cite{zhao2013tunable}, memristors \cite{Georgiou2018}, and microelectromechanical systems (MEMS) \cite{Kan2015}. 

Determining the characteristics of the unit cell that will lead to the desired behavior is typically performed analytically, through methods such as impedance matching \cite{PhysRevApplied.11.044024}. A large subset of designs has been based on the generalization of the Snell's laws of reflection and refraction \cite{Yu2011a, Moccia2017, Huang2017, Hosseininejad2019a}, which provides fundamental understanding on how to achieve certain functionalities through the drawing of specific phase gradients. When analytical methods are not practical, computational optimization methods \cite{Dong2015, Zhang2017b} or even machine learning approaches can be employed \cite{8889025, 8b03171, 5033327}. Another method worth remarking is that of coding metasurfaces, where the metasurface is \emph{encoded} using a discrete set of unit cell options or states \cite{Cui2014}. This allows drawing clear parallelism with information theory, enabling new ways to compose advanced metasurfaces by, for instance, adding or convoluting the codes of two desired \ac{EM} functionalities \cite{Cui2016, Liu2016b}. For instance, the design process is shown in Section \ref{sec:coding} follows the principles of generalized Snell's law and coding metasurfaces.

\begin{figure}[!t]
		\vspace{-0.7cm}
	\includegraphics[width=\columnwidth]{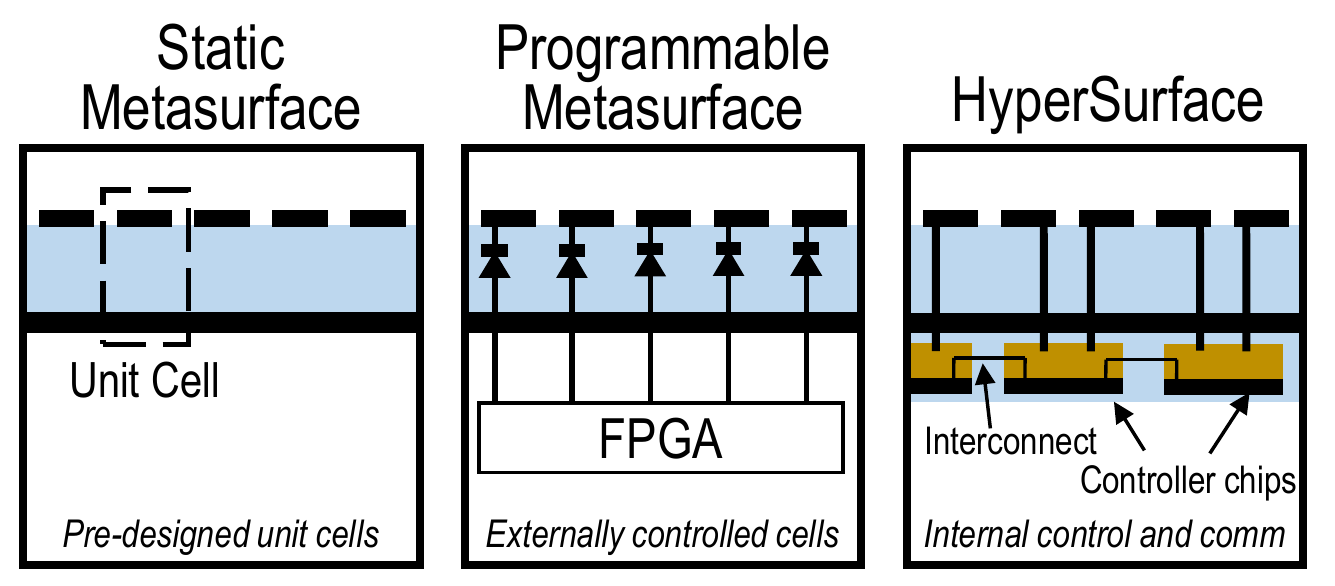} 

	\caption{Progression of metasurface design, from static to programmable with fully integrated tuning and control electronics.}
	\vspace{-0.4cm}
	\label{fig:background}
\end{figure}

The addition of control methods, together with the tunability on a per-cell basis, enables the programming of coding metasurfaces \cite{Liu2018ISCAS, Cui2014, Liu2017a}. The example in \cite{Yang2016} uses a Field-Programmable Gate Array (FPGA) to drive the pin diodes of a reconfigurable metasurface, where each pin diode determines the binary state of an individual unit cell. Such a device, however, exemplifies the potential reliability issues of programmable metasurfaces. In this particular case, the pin diodes enabling the reconfigurability may generate a considerable current density in the device, which could lead to electromigration. Also, the FPGA is a potential single point of failure, rendering the metasurface useless if the FPGA breaks down.

Beyond that, several authors have proposed to integrate a network of communicating chips within the metasurface containing actuators, control circuits, and even sensors \cite{AbadalACCESS, Tasolamprou2018, Liaskos2018a, PhysRevApplied.11.044024}. This concept, referred to as \ac{HSF}, opens new opportunities in the design of autonomous self-adaptive programmable metasurfaces but, at the same time, poses further challenges in the implementation, co-integration, and testing of the electronics within and around the metasurface.

\begin{figure}[!t]
	\subfigure[Front, back, and layered representation \cite{Pitilakis2018MMParadigm}]{\includegraphics[width=\columnwidth]{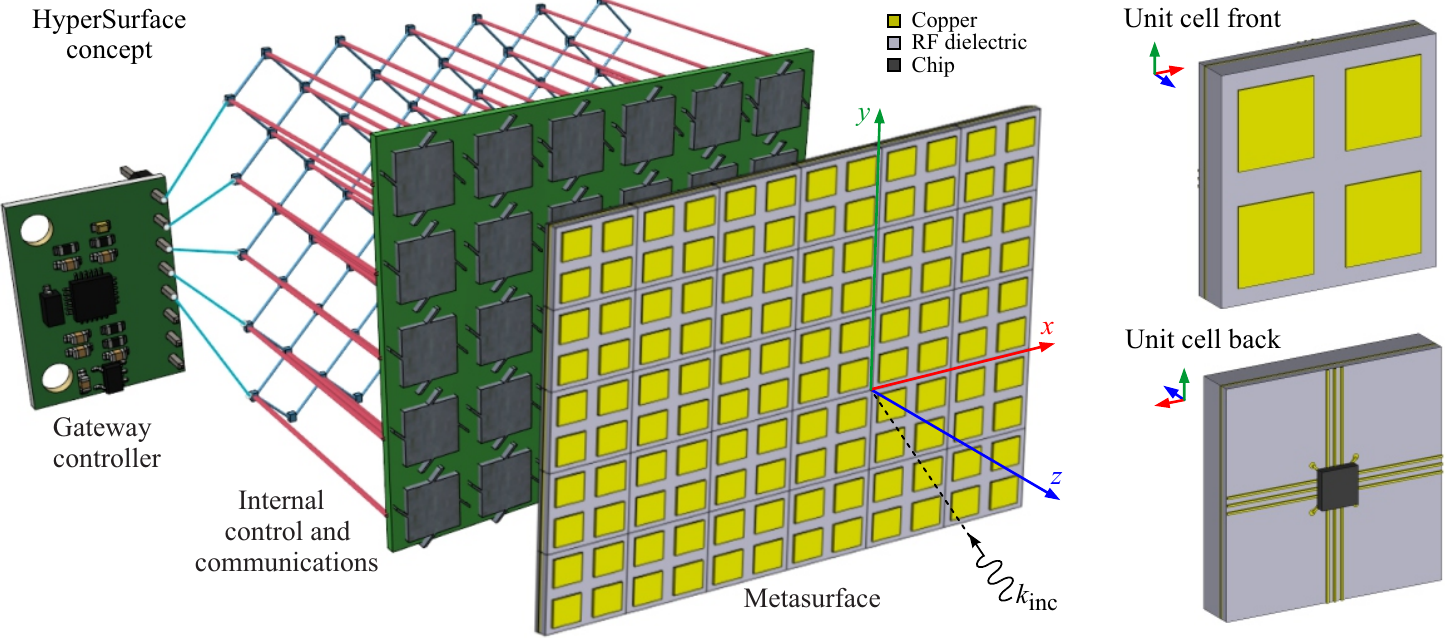}}
	\subfigure[Cross-section and logical planes]{\includegraphics[width=\columnwidth]{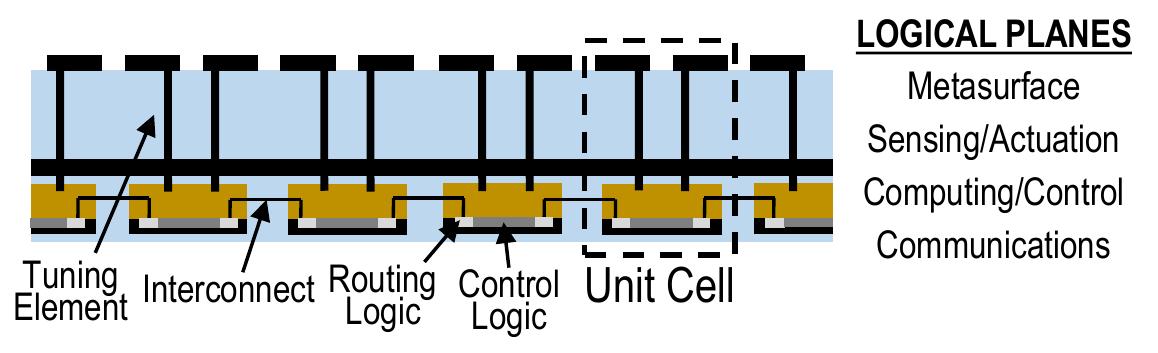}} 
	\vspace{-0.4cm}
	\caption{Graphical representation of a possible HSF implementation, which includes the metasurface plane with the metallic patches and the substrate, the sensing/actuation plane with the tuning elements and sensors, the computing/control plane containing the controller chips, and the communications plane containing the routing logic and interconnects. A gateway controller interfaces the HSF with the external world.}
	\vspace{-0.2cm}
	\label{fig:HSF}
\end{figure}

Figure \ref{fig:HSF} shows a graphical representation of the HSF structure, illustrating the logic planes of the device. Essentially, the HSF receives external programmatic commands from a gateway controller that are disseminated to the internal control logic at the controller chips via chip-to-chip interconnects and routing logic \cite{Saeed2018b, Saeed2019}. These commands contain the state (within the discrete set of possible states) that should be applied to each unit cell. The control logic translates the state into an analog value to be applied to the tuning element, e.g. the voltage applied to a varactor to achieve a target capacitance. Additionally, embedded sensors can pick up data from the environment and send it to the control logic or external devices again via the communications plane.

From the above, it seems clear that the integration of the controller chips, the chip-to-chip network, as well as the sensors and tuning elements, complicate the HSF design from the reliability perspective. As we will see in Section \ref{sec:error}, each of these components may fail in several ways leading to different consequences.

\section{Metasurface Model for Anomalous Reflection}
\label{sec:Sys}

In this paper, we evaluate the impact of faults in the beam steering capability of a programmable metasurface. To that end, the metasurface needs to implement anomalous reflection with the incidence and reflection angles as potential inputs. Here, we first describe the model used in this work to obtain the \ac{EM} response of each individual unit cell (Section \ref{sec:unit}) and of the complete metasurface in the far field (Section \ref{sec:meta}). Finally, we outline the methods to determine the coding of the metasurface, this is the states that need to be applied to each unit cell to direct the beam to a given target direction (Section \ref{sec:coding}).

\subsection{Unit Cell Model}
\label{sec:unit}
Unit cells are generally designed with a certain function in mind. For instance, the anomalous reflection will require unit cells to exhibit a reflection coefficient with high amplitude plus reconfigurable phase to change the angle of reflection \cite{Yu2011a}. As shown in several works, providing phase reconfigurability can be achieved via several tuning mechanisms \cite{PhysRevApplied.11.044024}. Since one of the aims of this paper is to capture the possible loss of performance arising from component faults, here we provide a particular unit cell design.

The case shown here revolves around the promising application of programmable metasurfaces in millimeter-wave communications for 5G (Fig. \ref{fig:intro}). We assume a square unit cell ($c=2$ mm) with a metallic backplane for operation around 25 GHz, aimed at giving service to one of the available $5G$ bands according to new recommendations by the International Telecommunication Union (ITU) \cite{itu2}. A square metallic patch ($b=1.85$ mm) is stacked on top of a substrate (Rogers RO4003C) with permittivity $\epsilon_r=3.5$ and thickness $a=0.81$ mm. It is possible to modify the phase response of the unit cell by adding capacitance to the square metallic patch. For phase tunability, this capacitance is given by varactors, which are embedded within the controllers and hidden under the backplane, but connected to the top patch with vertical vias \cite{Tasolamprou8788546}.

\begin{figure}[!ht]
			\vspace{-0.7cm}
	\includegraphics[width=1\columnwidth]{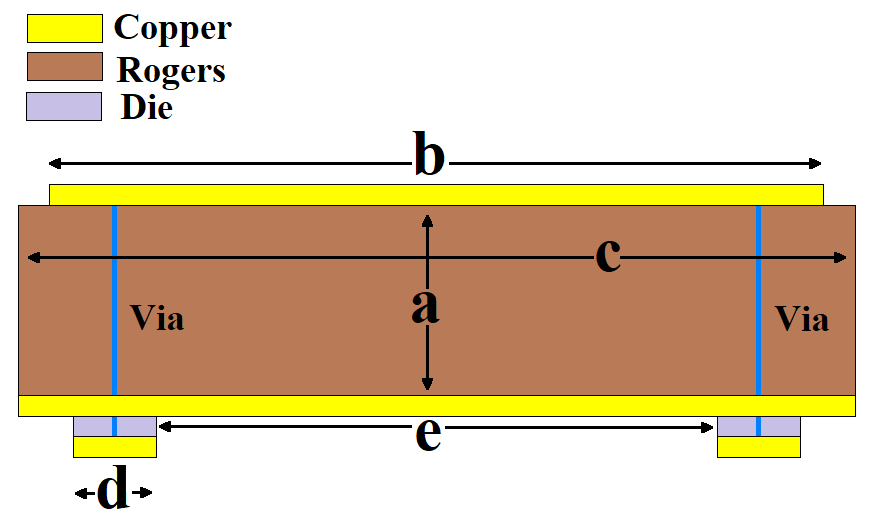} 
	\includegraphics[width=0.49\columnwidth]{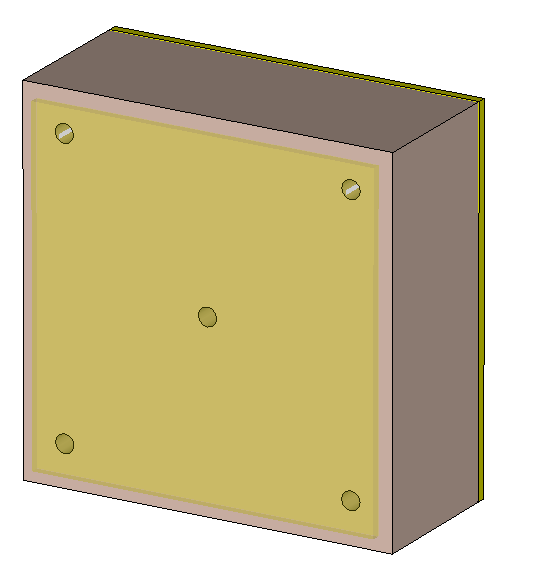} 
	\includegraphics[width=0.49\columnwidth]{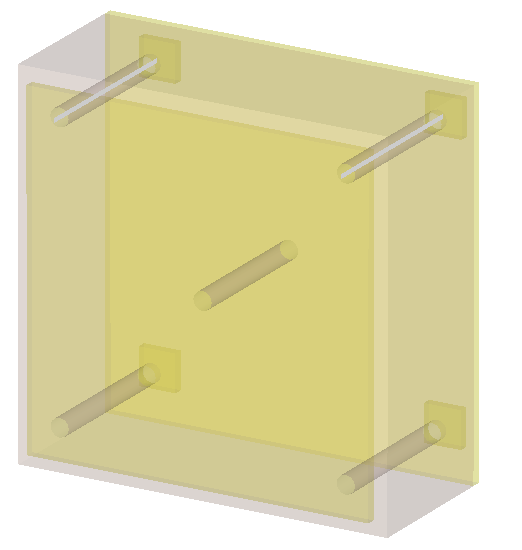} 
	\caption{Cross-section, top-view, and bottom-view of the assumed unit cell.}
		\vspace{-0.4cm}
	\label{Cell}
\end{figure}

We implement the proposed unit cell in a full-wave solver, CST MWS \cite{CST}, and evaluate the reflection coefficient when the unit cell is illuminated by a normal incident plane wave and for a set of capacitance values. 
 
Assuming that our design implements four coding states, it is standard practice in anomalous reflection metasurfaces that the $2\pi$ phase range is divided into evenly spaced states with $\pi/2$ separation with high reflection amplitude \cite{PhysRevApplied.11.044024}.  
 
\begin{figure}[!ht]
				\vspace{-0.2cm}
	\includegraphics[width=1\columnwidth]{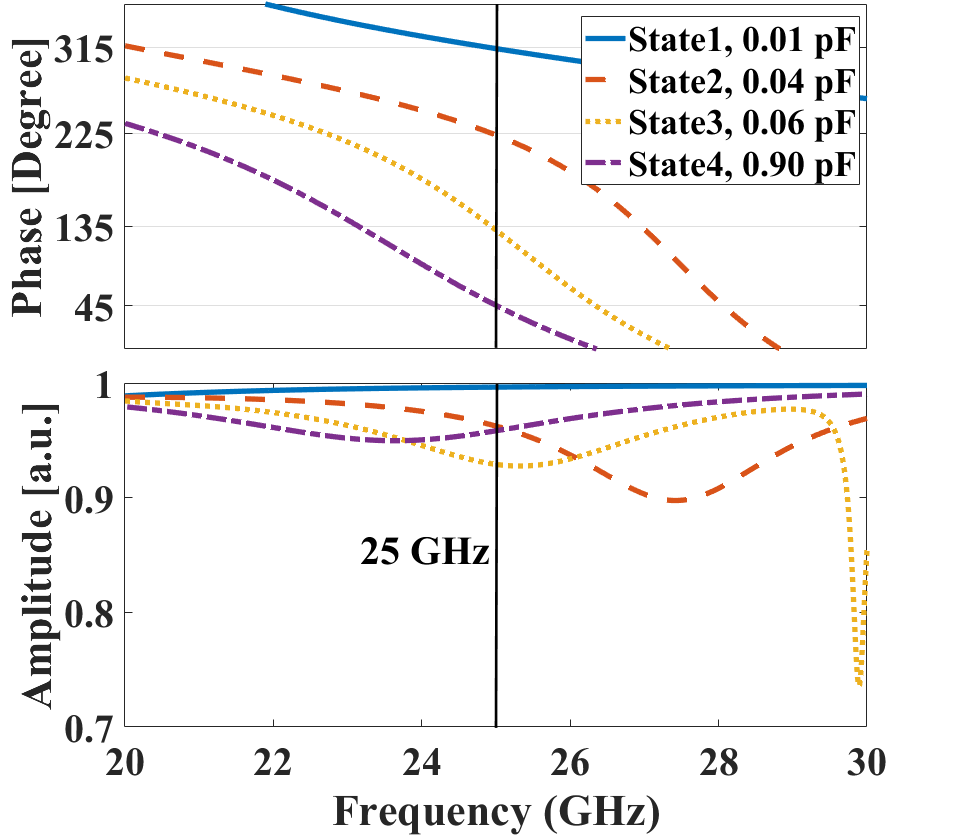} 
	\vspace{-.5cm}
	\caption{Unit cell reflection phase $\Phi$ (top) and amplitude $\Gamma$ (bottom) as a function of frequency for the four chosen capacitance values.}
	\label{Cellp}
\end{figure}

As shown in Figure \ref{Cellp}, the unit cell achieves these objectives around the target frequency, 25 GHz, with a reflection amplitude $\Gamma$ of 0.9 and phases $\Phi$ at $\{45, 135, 225, 315\}$ degrees. The figure plots the capacitances that have achieved such separation: 0.01 pF, 0.04 pF, 0.06 pF, and 0.9 pF. We will see that, if the capacitances deviate from such values, the unit cell may inaccurately point to different amplitude and phase.

				\vspace{-0.4cm}

\subsection{Metasurface Model}
\label{sec:meta}
Beam steering is a particular case of wavefront manipulation that occurs in the far field. As such, the metasurface can be accurately modeled as a compact array following the Huygens principle \cite{BalanisBOOK}. This method has been validated in several works via extensive simulations \cite{Hosseininejad2019a}. Considering each unit cell as an element of the array, the far field of the metasurface can be obtained as
\begin{equation}
F(\theta, \phi) = f_{E}(\theta, \phi) \cdot f_{A}(\theta, \phi) , 
\label{eq0}
\end{equation}
where $\theta$ is the elevation angle, $\phi$ is the azimuth angle of an arbitrary direction, $f_{E}(\theta, \phi)$ is the element factor (pattern function of unit cell) and $f_{A}(\theta, \phi)$ is the array factor (pattern function of unit cell arrangement). With the widespread assumption of a planar wave covering the entire metasurface, the scattering pattern will depend only on the array factor. For the metasurface with $N\times M$ unit cells, the far field pattern becomes 
\begin{equation}
\begin{split}
F(\theta, \phi) = \sum_{m=1}^{M} \sum_{n=1}^{N}A_{mn}e^{j\alpha_{mn}} f_{mn}(\theta_i, \phi_i)\\
\Gamma_{mn}e^{j\Phi_{mn}}f_{mn}(\theta, \phi)
e^{jk_0\zeta_{mn}(\theta, \phi)}
\end{split}
\label{eq1}
\end{equation}
where $A_{mn}$ and $\alpha_{mn}$ are the amplitude and phase of the wave incident to the $mn$-th unit cell; $\Gamma_{mn}$ and $\Phi_{mn}$ are the amplitude and phase of the response of the $mn$-th unit cell; $f_{mn}(\theta, \phi)$ denotes the scattering diagram of the $mn$-th unit cell towards an arbitrary direction of reflection, whereas $f_{mn}(\theta_i, \phi_i)$ denotes the response of the $mn$-th unit cell at the direction of incidence determined by $\theta_i, \phi_i$ and $k_0=2\pi/\lambda_{0}$ is the wave number (air is assumed). Finally, $\zeta_{mn}(\theta, \phi)$ is the relative phase shift of the unit cells with respect to the radiation pattern coordinates, given by
\begin{equation}
\zeta_{mn}(\theta, \phi) = D_u\sin{\theta}[(m-\tfrac{1}{2})\cos{\phi}+(n-\tfrac{1}{2}) \sin{\phi}]
\end{equation}

We further make the plausible assumption of plane wave incidence, so that factors $A_{mn}$, $\alpha_{mn}$, and $f_{mn}(\theta_i, \phi_i)$ are constants for all $m,n$. 

Further, we model the scattering pattern of the unit cell over the positive semisphere with the function $\cos(\theta)$, which is a widespread assumption \cite{Yang2016}. Finally, and without loss of generality, we consider the normal incidence ($\theta_i=\phi_i=0$). Then, Eq. \eqref{eq1} becomes
\begin{equation}
\label{eq2}
E(\theta, \phi) = K \cos\theta \sum_{m=1}^{M} \sum_{n=1}^{N} \Gamma_{mn} e^{j[\Phi_{mn}+k_0\zeta_{mn}(\theta, \phi)]}
\end{equation}
where $K$ is a constant. By controlling the phase shift of the unit cells $\Phi_{mn}$, we can implement anomalous reflection as described next.

\subsection{Metasurface Coding}
\label{sec:coding}
The coding of the cells of a programmable metasurface allows obtaining the desired functionality. 
In other words, we need to derive the amplitude $\Gamma_{mn}$ and phase $\Phi_{mn}$ of each unit cell so that the collective response matches with the required functionality. Then, we map the required $\Gamma$ and $\Phi$ to the closest available unit cell states.

In the case of anomalous reflection for beam steering, analytical methods provide high accuracy. Moreover, the error analysis performed in this work requires numerous iterations accounting for diverse combinations of faults --hence, the low computation times of analytical methods are desirable. In this work, we follow the well-known principles of wavefront manipulation whereby a phase gradient is used to determine the direction of reflection \cite{Yu2011a}. Assuming that the metasurface imposes the phase profile $\Phi(x,y)$, we assign the virtual wave vector $k_\Phi=\nabla \Phi_{x} \hat{x} + \nabla \Phi_{y} \hat{y} $ to the metasurface. In this context, the momentum conservation law for wave vectors can be expressed as
\begin{equation}\label{eq:dphi}
\begin{array}{l}
k_{i} \sin{\theta_{i}}\cos{\phi_{i}} + \frac{d\Phi_{x}}{dx} = k_{r} \sin{\theta_{r}}\cos{\phi_{r}} \\
k_{i} \sin{\theta_{i}}\sin{\phi_{i}} + \frac{d\Phi_{y}}{dy} = k_{r} \sin{\theta_{r}}\sin{\phi_{r}}
\end{array} 
\end{equation}
where $\tfrac{d\Phi_{x}}{dx}$ and $\tfrac{d\Phi_{y}}{dy}$ describe the gradients in the $x$ and $y$ directions, respectively.

Since we can address any given (oblique) wave with a translation formulation \cite{Liu2018b}, let us consider the normal incident wave case ($\theta_i=\phi_i=0$) without loss of generality. Assuming air as the medium of the incident and reflected wave, we can simplify the formulation above as
\begin{equation}
d\Phi_x=\frac{2\pi dx\cos\phi_r \sin\theta_r}{\lambda_0},\,\,d\Phi_y=\frac{2\pi dy \sin\phi_r \sin\theta_r}{\lambda_0}
\label{eq:dxdy}
\end{equation}
which express the change in phase ($\Phi_{x}$ and $\Phi_{y}$) that needs to be performed per unit of distance ($dx$ and $dy$) in the $x$ and $y$ directions. Then, we set the unit cell size ($d_x=d_y=D_u$) in Equation \eqref{eq:dxdy} to obtain the phase required at $mn$-th unit cell as
\begin{equation}
\Phi_{mn}=\frac{2\pi D_u (m\cos\phi_r \sin\theta_r+n\sin\phi_r \sin\theta_r)}{\lambda_0}
\end{equation}

To assign states to each unit cell, the required phase $\Phi_{mn}$ is calculated for all the unit cells. Then, a closest-neighbor mapping is done between the required phase and that provided by the different unit cell states. For instance, in our particular case where the number of states is $N_{s}=4$, we have $\{s_0,s_1,s_2,s_3\}$ with the respective phases of $\{45,135,225,315\}$ degrees, then required phases of 53\textsuperscript{o} and 188\textsuperscript{o} would be mapped to $s_0$ and $s_2$ states, respectively.

\section{Error Model}
\label{sec:error}

This section presents the model that we propose for the error analysis of programmable metasurfaces. The model describes both the impact of faults on the behavior of individual unit cells and how the faults can be distributed across the metasurface in Sections \ref{sec:types} and \ref{sec:spatial}, respectively. We also reason about the possible sources of each type of fault and attempt to exemplify a few relevant causes.

Generally speaking, faults in electronic systems may occur for a wide variety of reasons. This is also true in the mixed-signal HSF platform, where the metasurface and its associated tuning, control, and communication subsystems are integrated together. In any case, the relevance of different types of failure will eventually depend, among others, on the maturity of the technology, the manufacturing process, or the application environment. Several examples are outlined below.

For instance, it is widely known that chip failure rates and fabrication mismatches increase as the technology nodes go down \cite{Srinivasan2004}, which may become necessary in HSFs operating at mmWave and THz frequencies. Manufacturing defects could lead to stuck unit cells, similar to dead pixels in displays. When interconnecting the chips that drive the different unit cells, connector constraints or bad fitting can also lead to errors of different typologies. Once deployed, chip connections might fail over time due to thermal cycling or flexing. Metasurfaces could be exposed to physically challenging conditions that could lead to hard faults, such as physical damage in a conflict zone where bullets could impact the metasurface or bit flips due to cosmic radiation in space applications. Last but not least, ultra-low-power HSFs could power-gate a set of controllers in order to save energy in environments where a given performance degradation is tolerable. Here, the error analysis would help to determine which controllers should be powered off and at which state they should be kept. In any case, power gating can be regarded as an intentional transient fault.

\subsection{Types of Errors}
\label{sec:types}
Here, we describe the impact that faults can have on the performance of individual unit cells. We assume that each unit cell is assigned a \emph{valid state} $s \in \Sigma$, where $\Sigma$ represents the set with cardinality $N_{s}$ of valid states for a particular unit cell design. As shown in Section \ref{sec:Sys}, the state $s$ basically determines the amplitude and phase of the reflection coefficient at an arbitrary unit cell. 

\begin{figure*}[!h]
				\vspace{-0.7cm}
	\includegraphics[width=0.5\columnwidth]{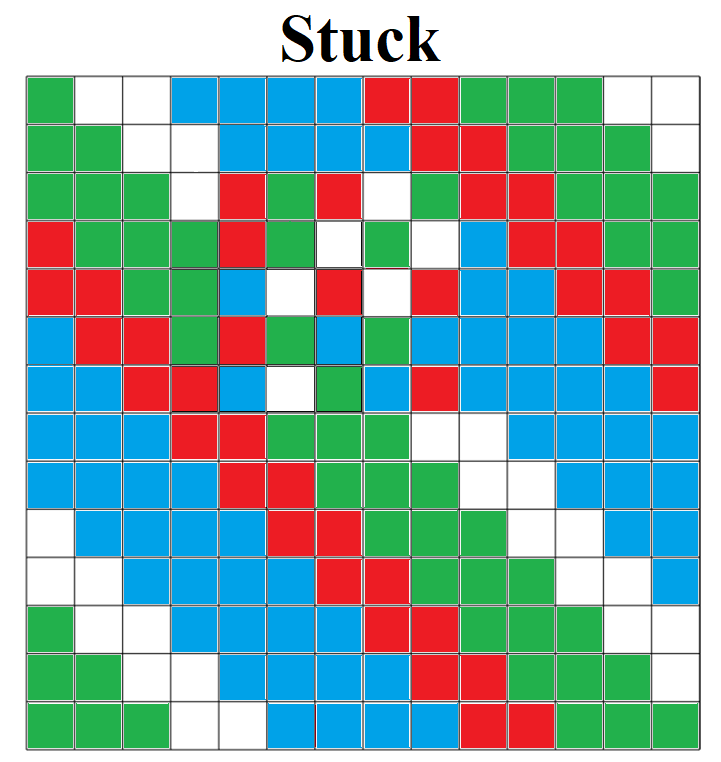} 
	\includegraphics[width=0.5\columnwidth]{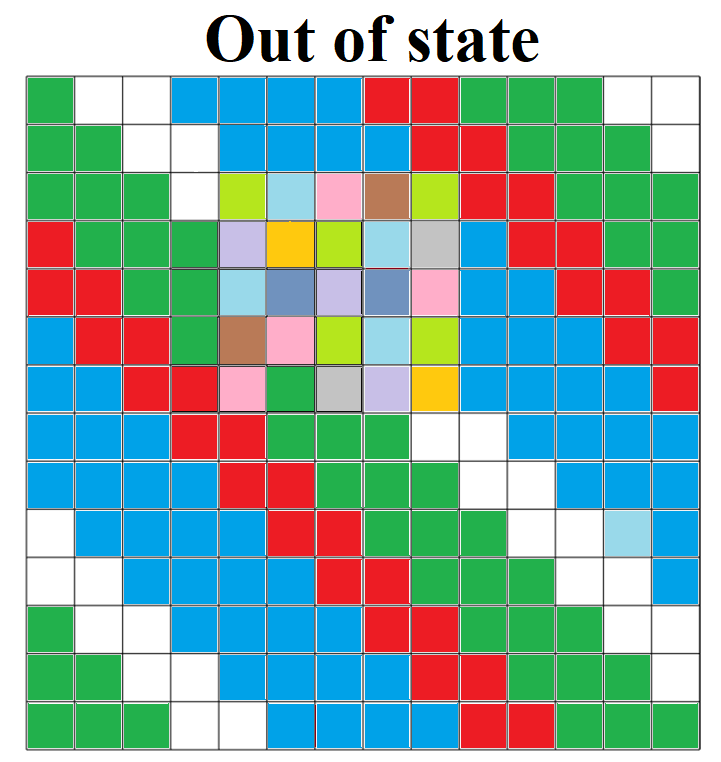}
	\includegraphics[width=0.5\columnwidth]{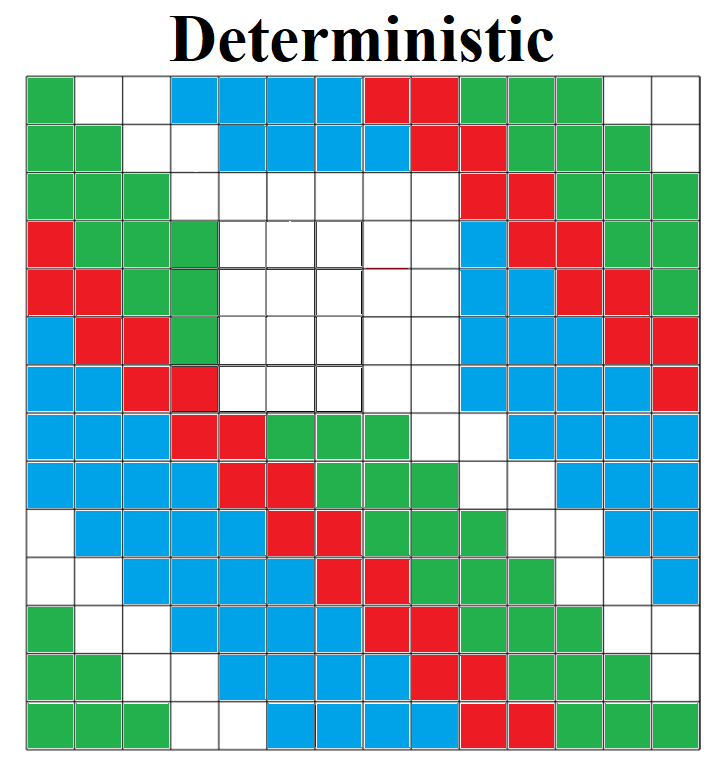} 
	\includegraphics[width=0.5\columnwidth]{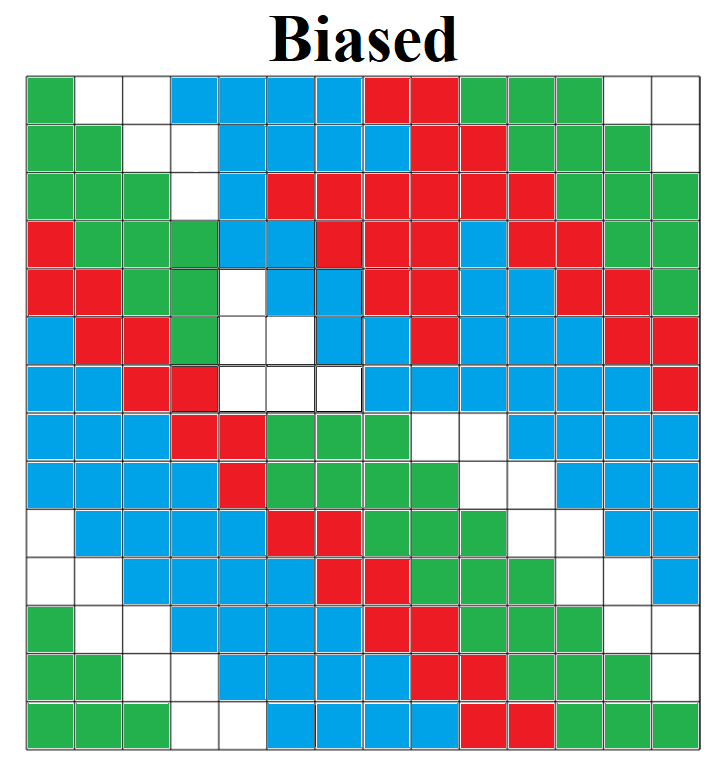} 
	\caption{Graphical representation of the different types of error in a metasurface with 15$\times$15 unit cells. White, green, red, and blue squares indicate valid states, whereas other colors indicate invalid states.}
				\vspace{-0.4cm}
	\label{fig:type}
\end{figure*}

In the presence of a fault, we assume that the unit cell will transition to a state $s'$ which may or may not be within the set of valid states of the metasurface. The value of $s'$ and its probability will depend on the type of fault, that we comprehensively classify as follows (see Figure \ref{fig:type}):

\begin{itemize}
	
	\item \textbf{Stuck at state:} the unit cell is stuck at a random, but valid unit cell state $s'\in \Sigma$. This type of error assumes that a failure \emph{disconnects} the unit cell from the rest of the system, leaving it in an old state. Such a disconnection can occur due to failures in the communication or control planes (e.g. in the router or in the controller) that prevent control signals to reach the tuning element. The value of $s'$ is picked randomly with uniform distribution.

	\item \textbf{Out of state:} the unit cell takes a random invalid state $s'\notin \Sigma$, which essentially means random amplitude and/or phase. Possible causes of this error may be failures that affect the tuning elements and, thus, lead to a wrong capacitance. Via disconnections arising from manufacturing defects or aging, or defects in the DAC circuits that drive the tuning element, could lead to such error.

	\item \textbf{Deterministic:} The unit cell stays in a known fixed, generally invalid state, which is the same across all unit cells with the same type of fault. For instance, a deterministic error could be caused by a physically damaged unit cell, i.e. a bullet making a small hole within the HSF could be approximated as zero phase and full transmittance.
	
	\item \textbf{Biased:} The unit cell is at a state which is at a fixed given distance $\Delta$ of the actual required state $s' = s+\Delta \in \Sigma$. This may be caused by flip-bit errors at the computing plane, or by external biases, perhaps due to attacks.
	
\end{itemize}

\hl{It is worth discriminating the mapping sequence from types of errors. In the end of Section {\ref{sec:coding}}, a process is described which maps the desired phase (in the $[0^o,360^o]$ range) to the closest available phase for four states $s_0,s_1,s_2,s_3$. Errors occur after the mapping done by design and, thus, may cause the unit cells states to shift to valid states (e.g. $s_0,s_1,s_2,s_3$; yet not the intended state for that unit cell) or invalid states (outside the phases mapped to $s_0,s_1,s_2,s_3$).}

Next, we illustrate how single component faults can affect the performance of an individual unit cell. In particular, we evaluate the impact of biasing a single varactor to the wrong voltage or completely disconnecting it. To that end, we simulate the different combinations using our unit cell design from Section \ref{sec:unit} as a baseline and calculate the phase error as the difference between the correct and erroneous reflection phase.

\begin{figure}[!ht]
				\vspace{-0.2cm}
	\includegraphics[width=1\columnwidth]{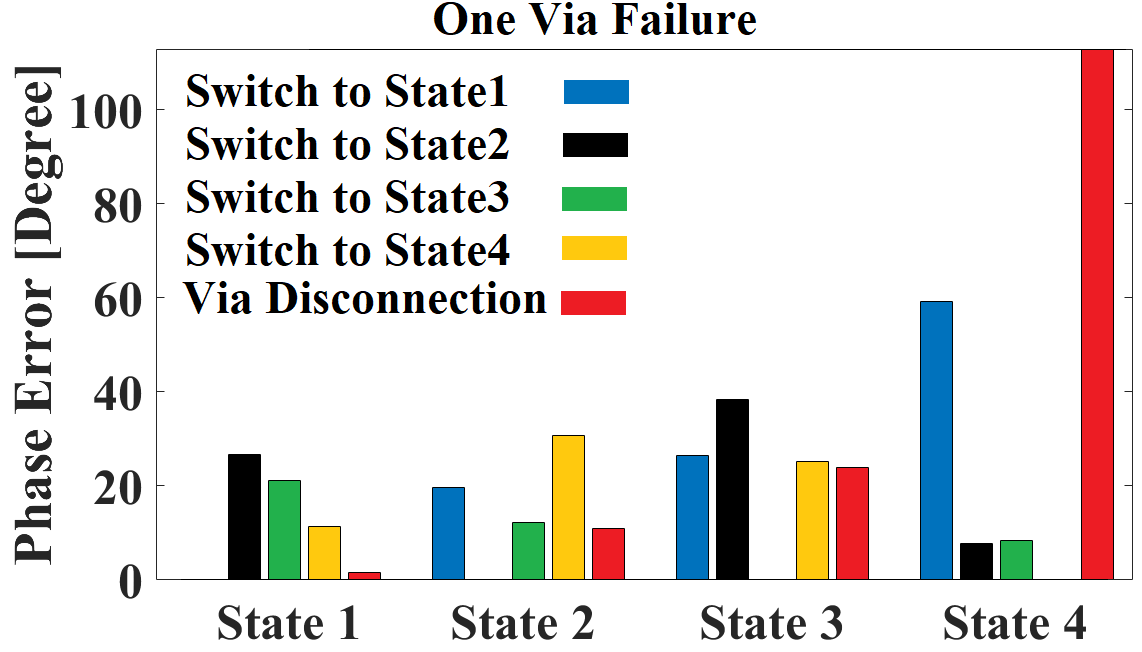} 
	\caption{Phase error resulting from the biasing of a single Via to an incorrect state, or its complete disconnection, for the proposed unit cell design at 26 GHz. The X-axis depicts the state at which the other Vias are biased.}
				\vspace{-0.4cm}
	\label{viaf}
\end{figure}

The results of this example are shown in Figure \ref{viaf}. Red bars represent the impact of Via disconnection with respect to the initial states. Exponential growth from state 1 (0.01 pF) to state 4 (0.9 pF) is observed, revealing that the disconnection of large capacitances has a more significant impact. Blue, black, green and yellow bars indicate the phase error resulting from the wrong biasing of a single Via from the state indicated in the X-axis to state 1, 2, 3, and 4 respectively. For instance, the first black bar indicates the phase error of biasing the Via to state 2 instead of state 1. The impact of each change is subtle and does not follow a clear trend. In this particular example, then, we could approximate single Via failures as an \emph{out of state} error with a random phase.

\subsection{Spatial Distribution}
\label{sec:spatial}
\begin{figure*}[!h]
				\vspace{-0.7cm}
	\includegraphics[width=0.5\columnwidth]{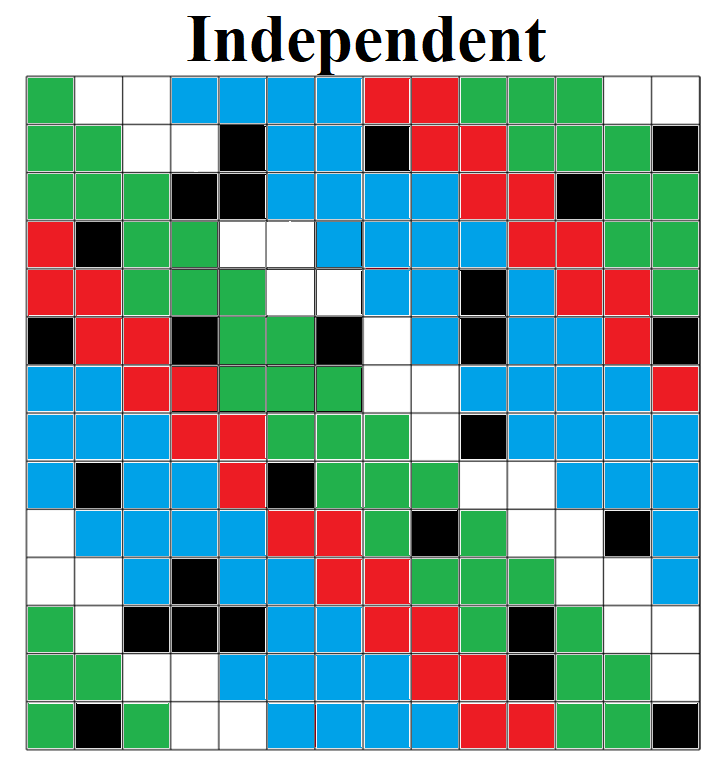}
	\includegraphics[width=0.5\columnwidth]{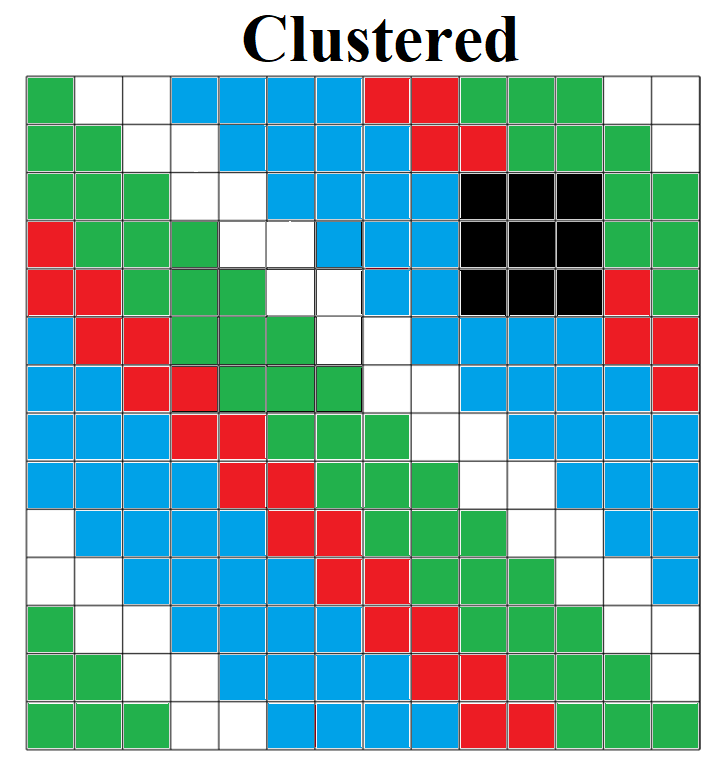} 
	\includegraphics[width=0.5\columnwidth]{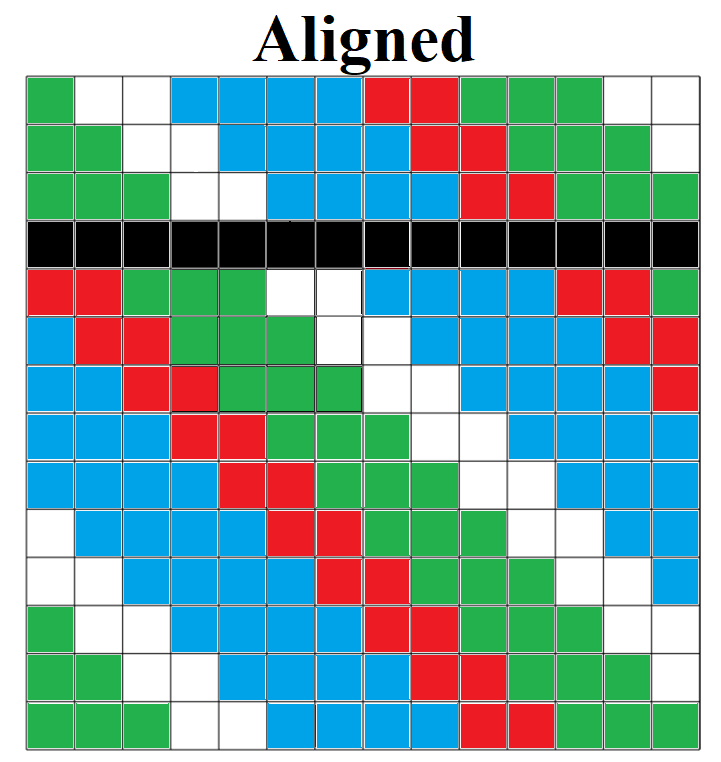}
	\includegraphics[width=0.5\columnwidth]{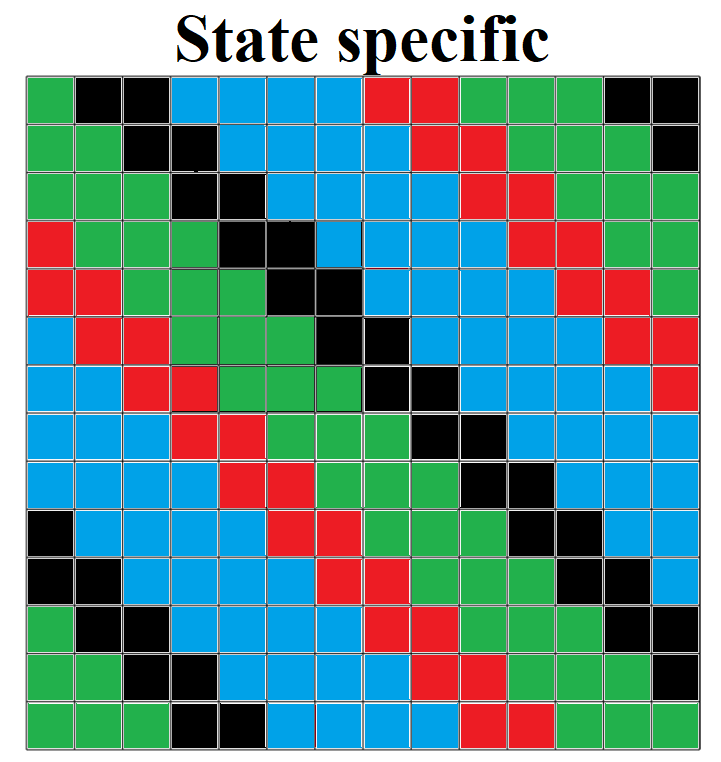}
	\caption{Graphical representation of the different error distributions in a metasurface with 15$\times$15 unit cells. Black squares represent faulty unit cells.}
				\vspace{-0.4cm}
	\label{fig:dis}
\end{figure*}

Next, we describe the possible spatial distribution of errors across the metasurface. We base our reasoning on the fact that faults may impact multiple unit cells or have cascading effects, this is, lead to further faults. We distinguish between the following distributions, represented in Figure \ref{fig:dis}:

\begin{itemize}
	\item \textbf{Independent:} The errors are randomly distributed over the metasurface and can be modeled with a spatial Poisson process. Individual uncorrelated faults, maybe with different origins, could yield such a distribution.
	\item \textbf{Clustered:} The errors appear around a given area. Cascading effects of a fault or faults that affect several unit cells can lead to such behavior. For instance, faults rendering a controller chip useless will impact all its associated unit cells. Another example would relate to the loss of connectivity at the network: faults in a few interconnects can leave an entire region of the metasurface isolated and stuck in an old state \cite{Saeed2018b}. 
	\item \textbf{Aligned:} The errors are spatially co-located following a line. For instance, let us assume that power or ground signals are distributed through the HSF through a matrix of electrical lines. We speculate that, in such a case, if one line representing a row or column fails, the whole row and column could be affected.
	\item \textbf{State-specific:} Another speculative type of spatial distribution would be that all unit cells that are supposed to be in a specific state, behave incorrectly. This could happen if the actuator uses an external value (e.g. voltage from a centralized regulator) to determine the given state; if that value is incorrect, the state will be erroneous.
\end{itemize}

\section{Methodology}
\label{sec:methodology}

A general methodology for the analysis of errors in metasurfaces would simply evaluate the metasurface in the presence of an increasing number of errors, and compare it with the performance of a golden reference. 

Basically, the golden reference is coded according to the \ac{EM} functionality and evaluated using the methods exemplified in Section \ref{sec:Sys} for beam steering. Then, faults are introduced using the guidelines discussed in Section \ref{sec:introducing}. To better understand the impact of errors, a set of representative metrics is used to characterize the performance degradation as discussed in Section \ref{sec:metrics}.

\subsection{Introducing Errors}
\label{sec:introducing}
\begin{figure}[!h]
	\centering
				\vspace{-0.2cm}
	\includegraphics[width=.7\columnwidth]{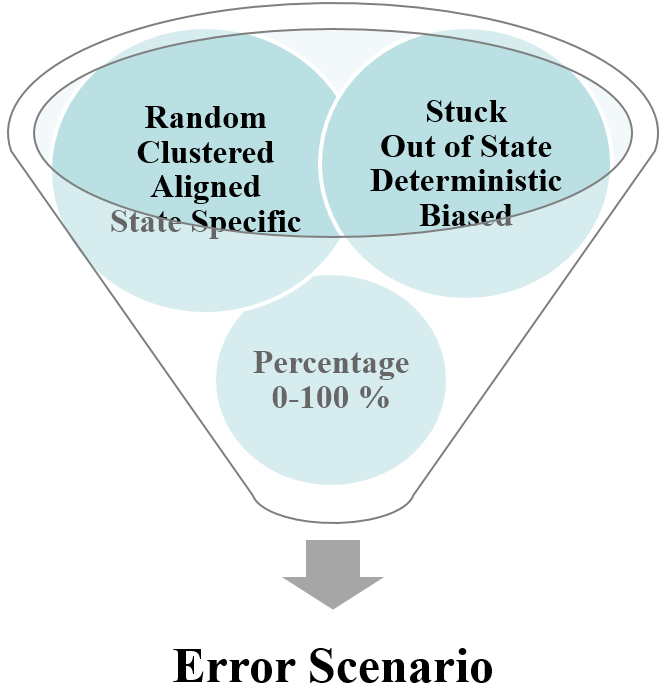} 
	\caption{Error scenario generation process.}

	\label{fig:sce}
\end{figure}

The analytical formulation allows to trivially introduce errors by modifying the terms $\Gamma_{mn}$ and $\Phi_{mn}$ of the affected unit cells in Equation \eqref{eq2}. The type of error, its spatial distribution, together with the percentage of faulty unit cells, define the error scenario as represented in Figure \ref{fig:sce}. To apply a particular error scenario, the steps are:
\begin{enumerate}
	\item To set the number of faulty unit cells according to the input percentage.
	\item To set the position of the faulty unit cells according to the spatial distribution, setting the $m$ and $n$ values in $\Gamma_{mn}$ and $\Phi_{mn}$, using spatial Poisson processes if required.
	\item To set the $\Gamma$ and $\Phi$ values of each particular unit cell, either within the discrete set of valid states $\Sigma$ or within a continuous range (valid or invalid states), depending to the type of error and using random number generators if required.
\end{enumerate}

To assess the validity of the analytical model, the far field obtained in a particular case is compared with that of a full-wave simulation performed with CST Microwave Studio \cite{CST}. 

\begin{figure}[!h]
				\vspace{-0.2cm}
	\centering
	\includegraphics[width=1\columnwidth]{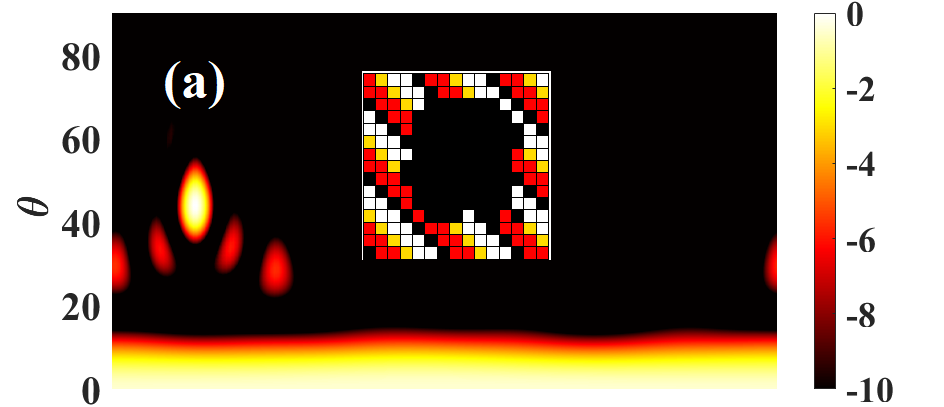} 
	\includegraphics[width=1\columnwidth]{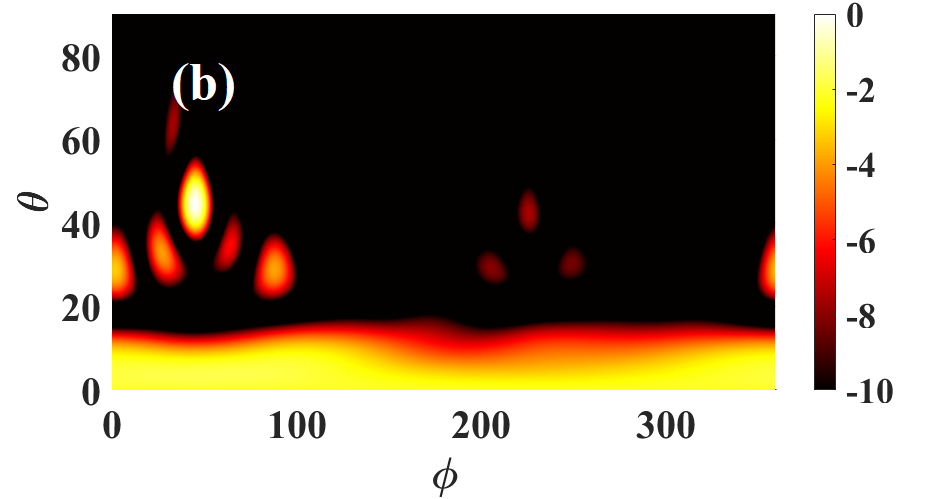} 
	\caption{Far field pattern, in dB, of a metasurface steering the beam towards $\theta_r=\phi_r=\pi/4$ with 30\% of faulty unit cells with deterministic,  clustered characteristics, obtained (a) analytically and (b) through full-wave simulation.}
				\vspace{-0.4cm}
	\label{fig:eval}
\end{figure}

The metasurface contains 15 unit cells per dimension. As shown in Figure \ref{fig:eval}, the analytical method is in close agreement with the simulation. It is worth noting that while the simulation takes 30--40 minutes in a high-end workstation, the analytical method calculates the results within several seconds.

\subsection{Performance Metrics}
\label{sec:metrics}
Far field analysis can provide valuable, yet mainly qualitative information about the impact of errors. To quantify the performance degradation and extract behavioral trends, the use of a set of performance metrics is suggested.

\begin{figure}[!h]
	\centering
		\vspace{-0.4cm}
	\includegraphics[width=.6\columnwidth]{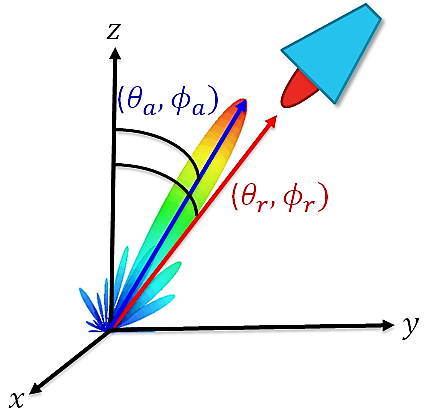}
	\caption{Discrimination between the target angle $(\theta_r, \phi_r)$ and the actual reflected angle $(\theta_a, \phi_a)$.}
	\label{DTD}
	\vspace{-0.4cm}
\end{figure}

In this work, we consider the following performance metrics related to beam steering:
\begin{itemize}
	\item \textbf{Target deviation ($TD$):} The main purpose of beam steering is to reflect the EM wave toward a determined angle. However, there is usually a difference between the targeted angle and the actual reflected angle due to inaccuracies in the phase profile or, in our case, the appearance of errors. We call this factor target deviation and calculate it as the Euclidean distance between the angle pointing to the direction of the reflected main lobe ($\theta_a,\phi_a$) and the desired angle for steering ($\theta_r, \phi_r$) as
	\begin{equation}
	TD=\sqrt{(\theta_r - \theta_a)^2+(\phi_r - \phi_a)^2}.
	\end{equation}
	
	\item \textbf{Directivity ($D(\theta, \phi)$):} As a fundamental antenna parameter, the directivity describes concentration of energy at a given direction. We evaluate it in specific relevant angles such as the target angle of reflection $D(\theta_r, \phi_r)$ and the actual angle of reflection $D(\theta_a, \phi_a)$ (see Figure \ref{DTD}).
	
	\item \textbf{Secondary lobe level ($SLL$):} The $SLL$ is defined as the ratio (in dB) of the far field strength in the direction of the side-lobe nearest to the main beam to the far field strength of the main beam. 
	
	\item \textbf{Side Lobe Accumulation ($SLA$):} In addition to the secondary beam, a set of minor reflected beams may appear due to the fundamental operation of MS structure. We measure the accumulation of power within these lobes and report it normalized to the power of the main beam.
	
	\item \textbf{Half power beam width ($HPBW$):} The waist of the main reflected beam defines the resolution of steering. The HPBW, that calculates the beam width at the -3dB of a normalized lobe, is a conventional factor to assess the beam width. 
	
\end{itemize}

\section{Results}
\label{sec:results}
This section applies the proposed methodology on a particular case of beam steering metasurface. We take, as the basic building block, the unit cell described in Section \ref{sec:unit} with the four states represented in Figure \ref{Cellp}.

We consider a metasurface of 15$\times$15 unit cells, coded to target $\theta_r=\phi_r=\pi/4$ from normal incidence, using the methods described in Section \ref{sec:coding}. The far field is obtained with equations from Section \ref{sec:meta}. Indexing the performance metrics, we get $D(\theta_r,\phi_r)=0$ dB, $D(\theta_a,\phi_a)=0$ dB, $TD=1.5$\textsuperscript{o}, $HPBW=14.47$\textsuperscript{o}, $SLL=-11.43$ dB and $SLA=11.14$ dB. Note that the directivity values are normalized to the strength at the direction of maximum radiation, which is why we obtain a value of 0 dB.

\subsection{Overview}
\begin{figure*}[!ht]
	\centering
				\vspace{-0.7cm}
	\subfigure[Sketch of the metasurface coding under Clustered-Stuck errors and radiation pattern for three error percentages.\label{CS}]
	{\includegraphics[width=0.5\columnwidth]{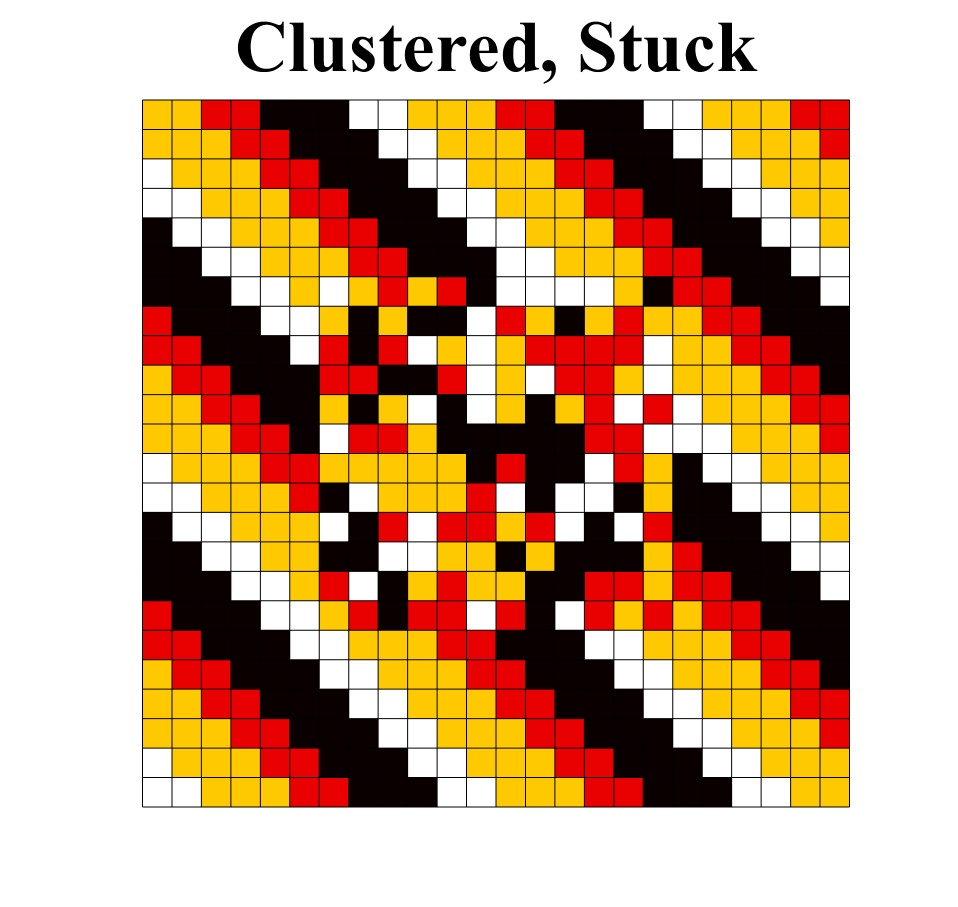}
		\includegraphics[width=0.5\columnwidth]{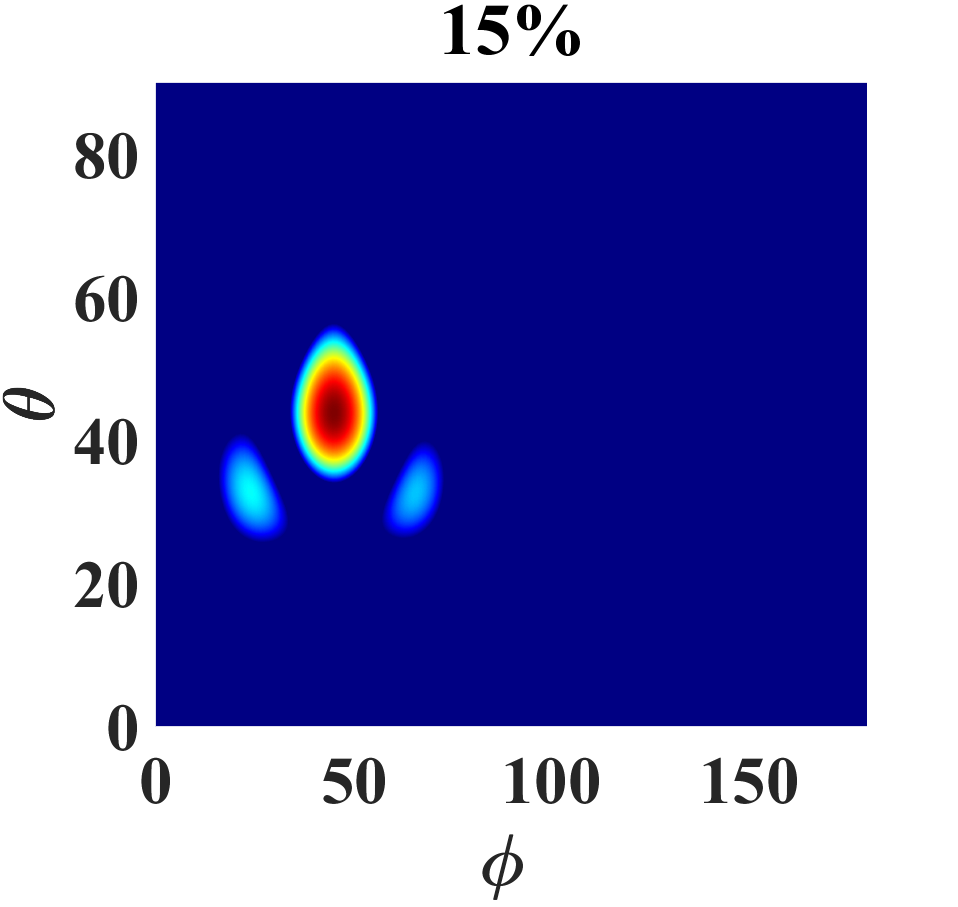}
		\includegraphics[width=0.5\columnwidth]{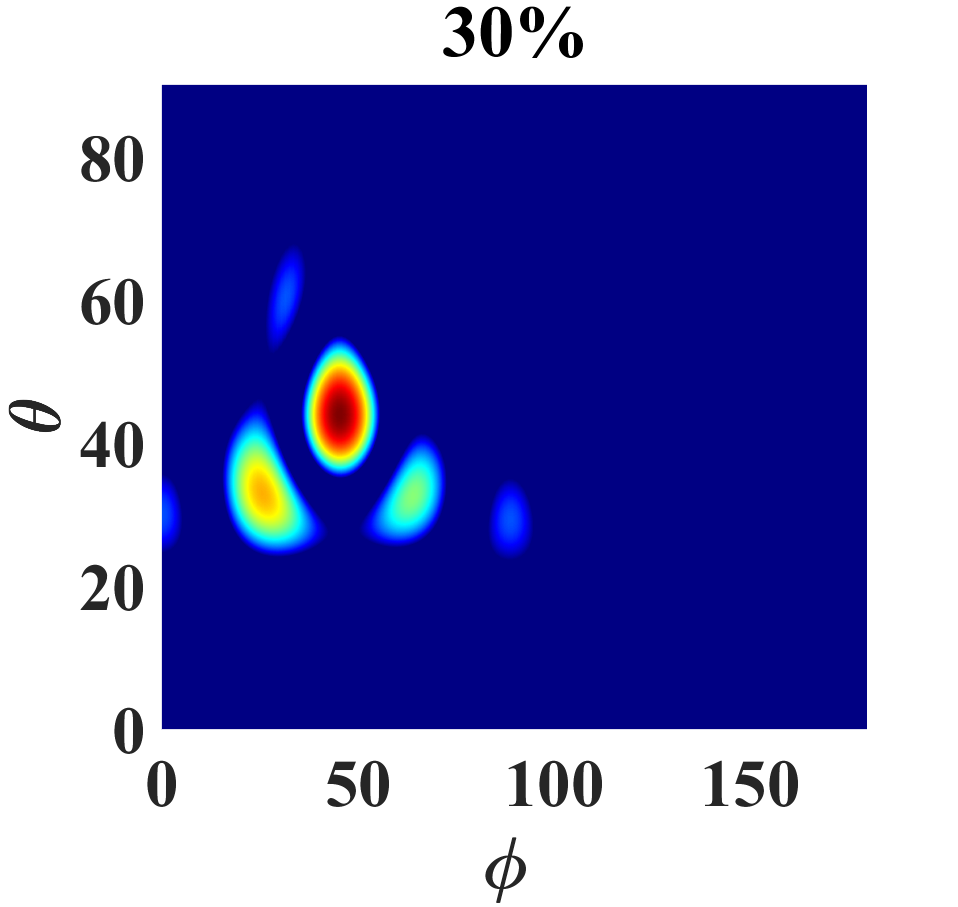}
		\includegraphics[width=0.5\columnwidth]{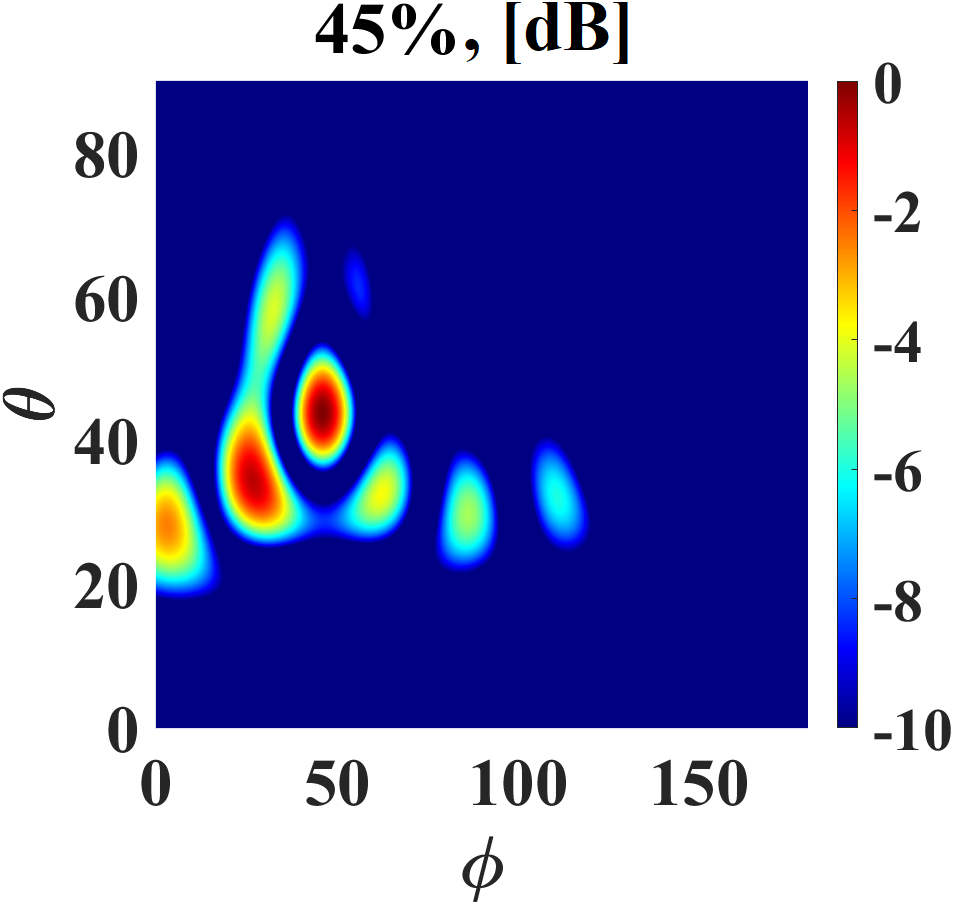}}

	\subfigure[Sketch of the metasurface coding under Clustered-Biased errors and radiation pattern for three error percentages.\label{CB}]
	{\includegraphics[width=0.5\columnwidth]{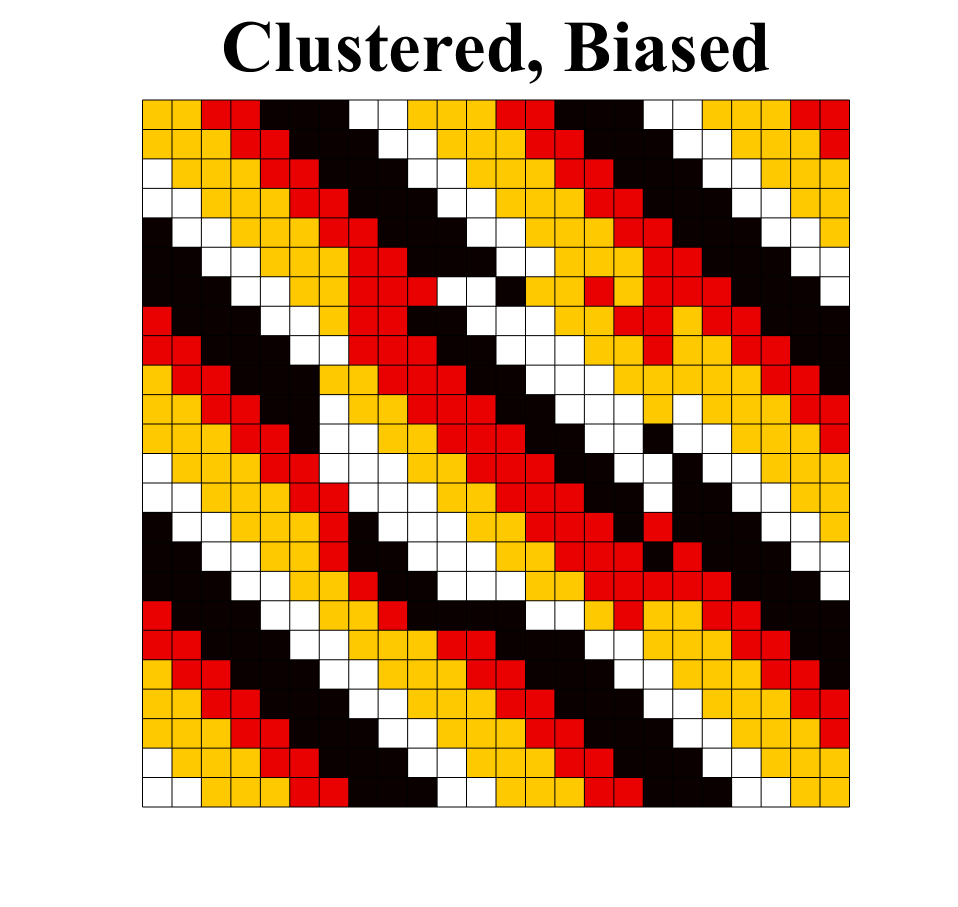}
		\includegraphics[width=0.5\columnwidth]{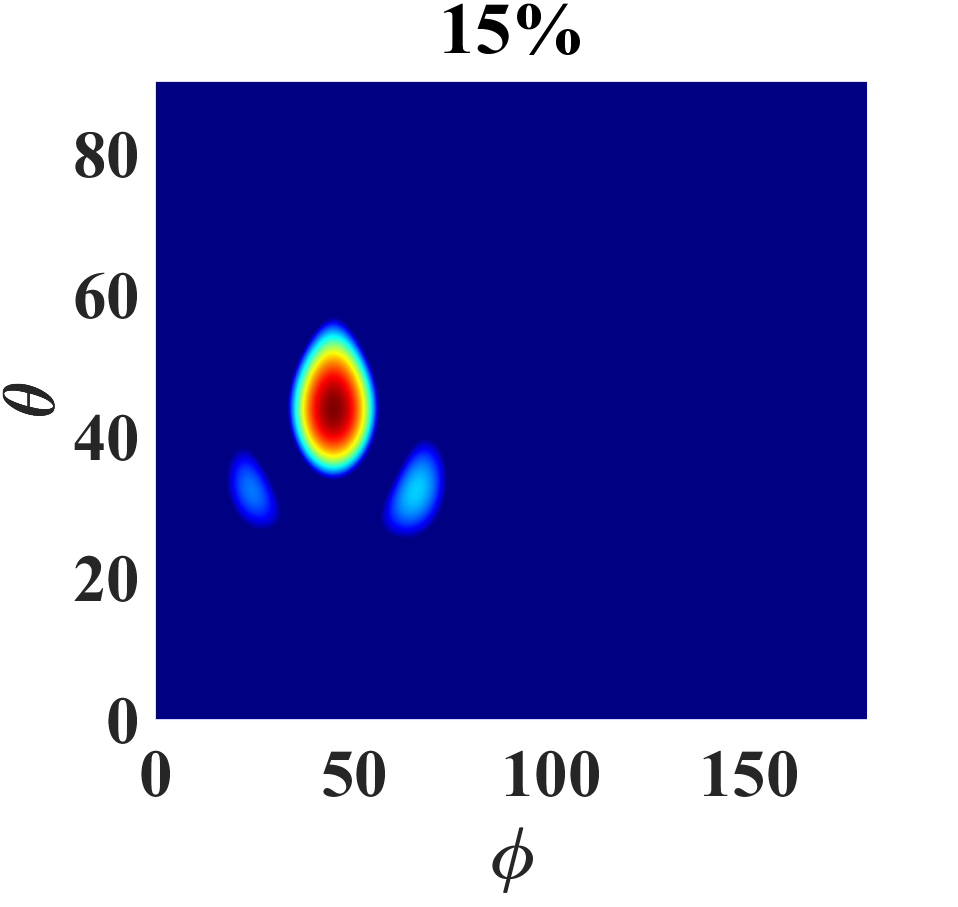}
		\includegraphics[width=0.5\columnwidth]{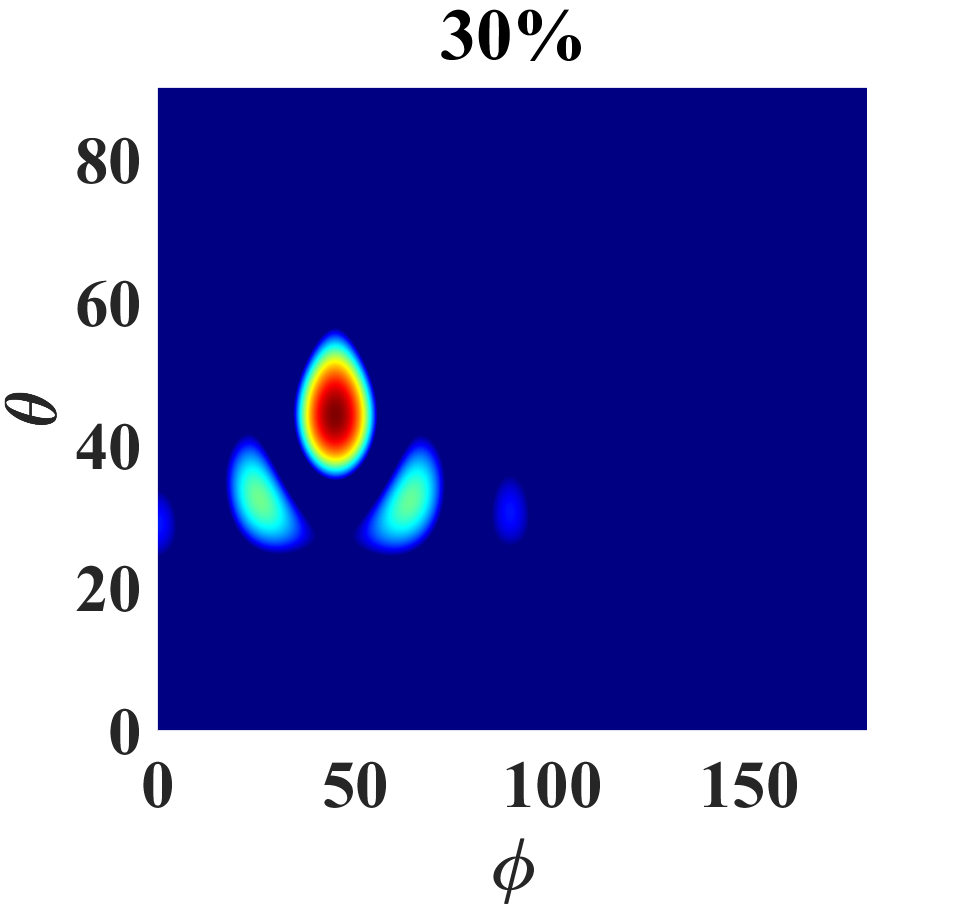}
		\includegraphics[width=0.5\columnwidth]{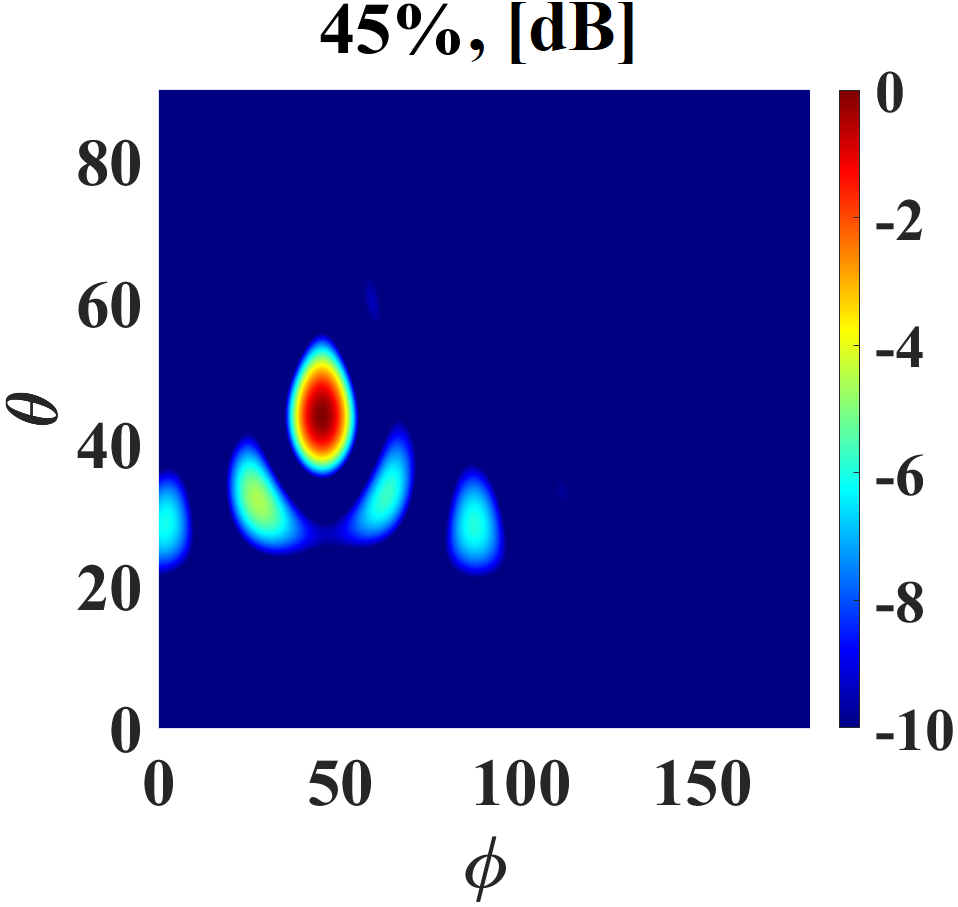}}
	\subfigure[Sketch of the metasurface coding under Independent-Out of state errors and radiation pattern for three error percentages.\label{IO}]
	{\includegraphics[width=0.5\columnwidth]{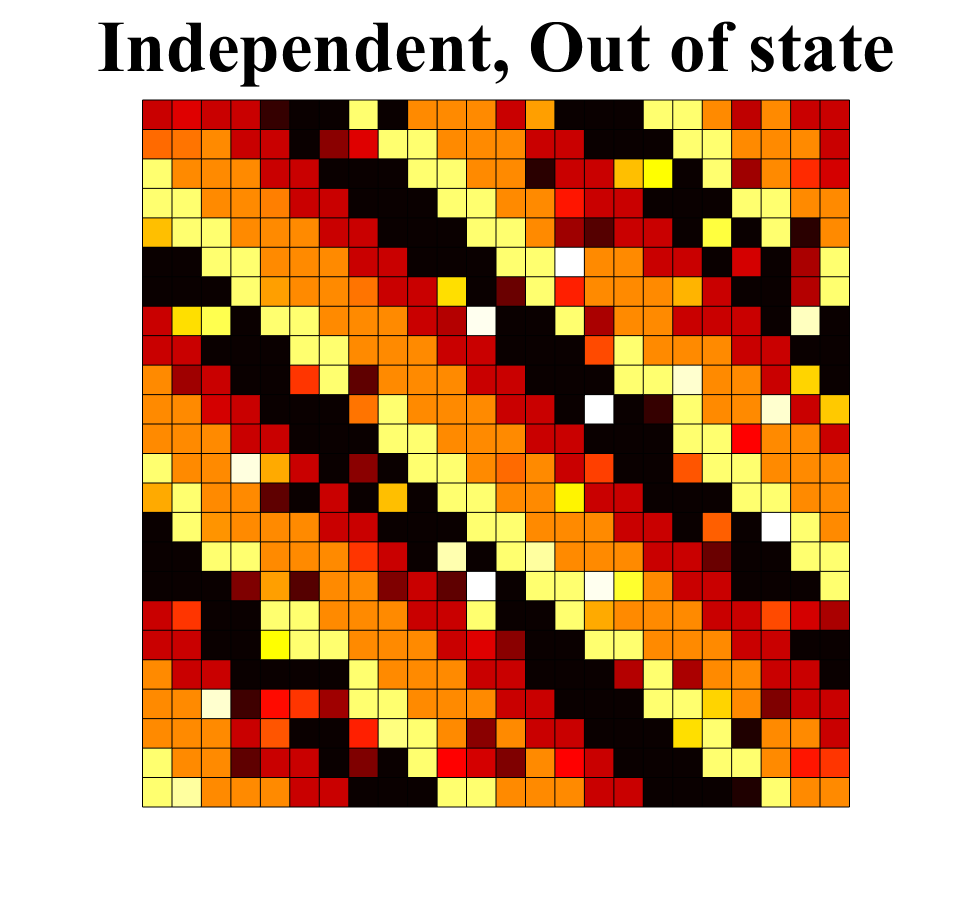}
		\includegraphics[width=0.5\columnwidth]{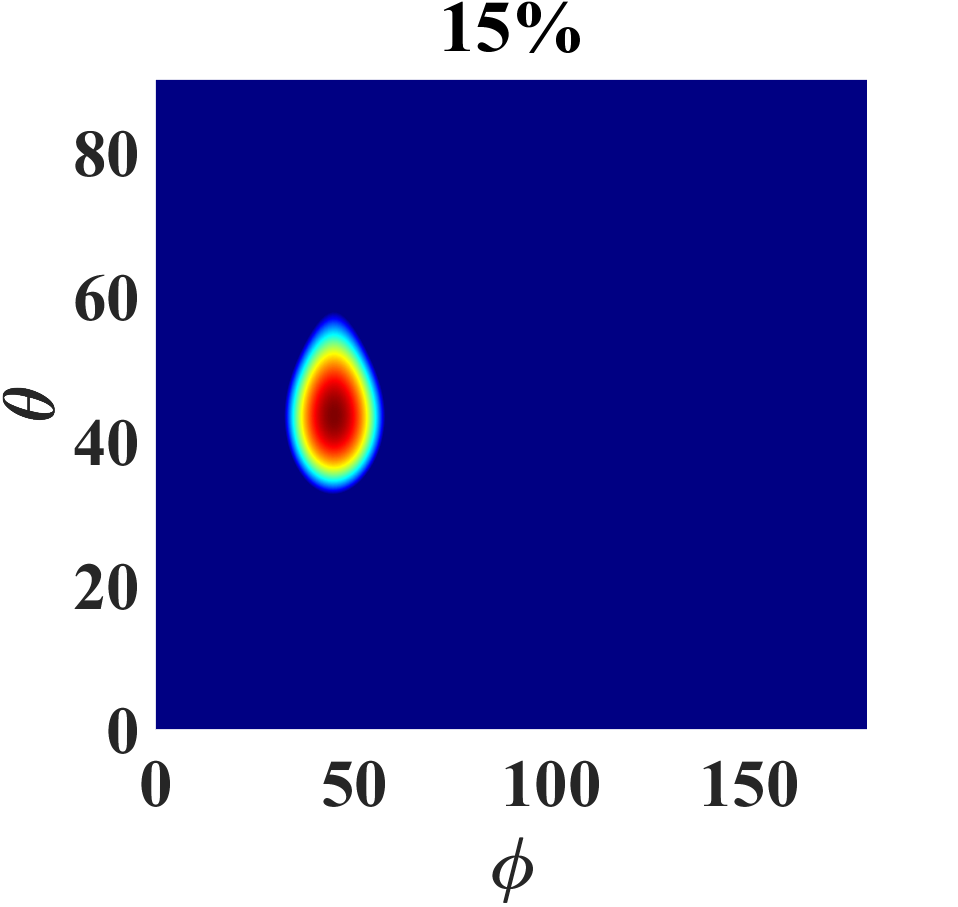}
		\includegraphics[width=0.5\columnwidth]{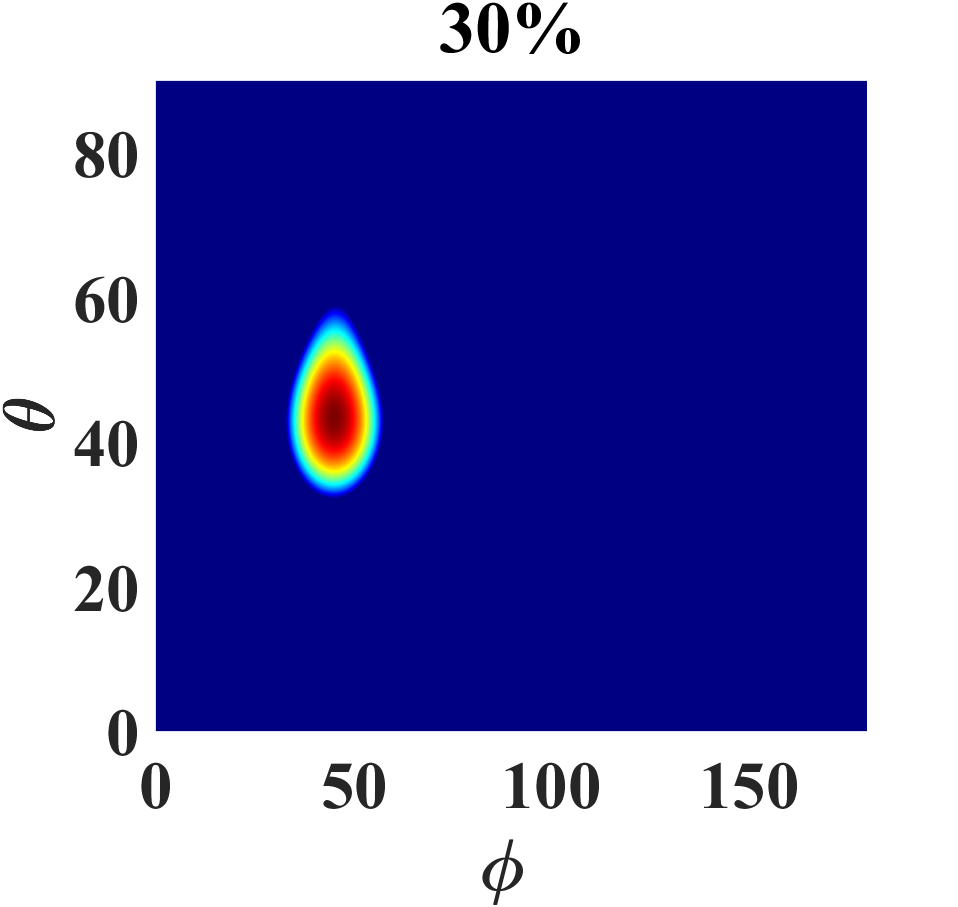}
		\includegraphics[width=0.5\columnwidth]{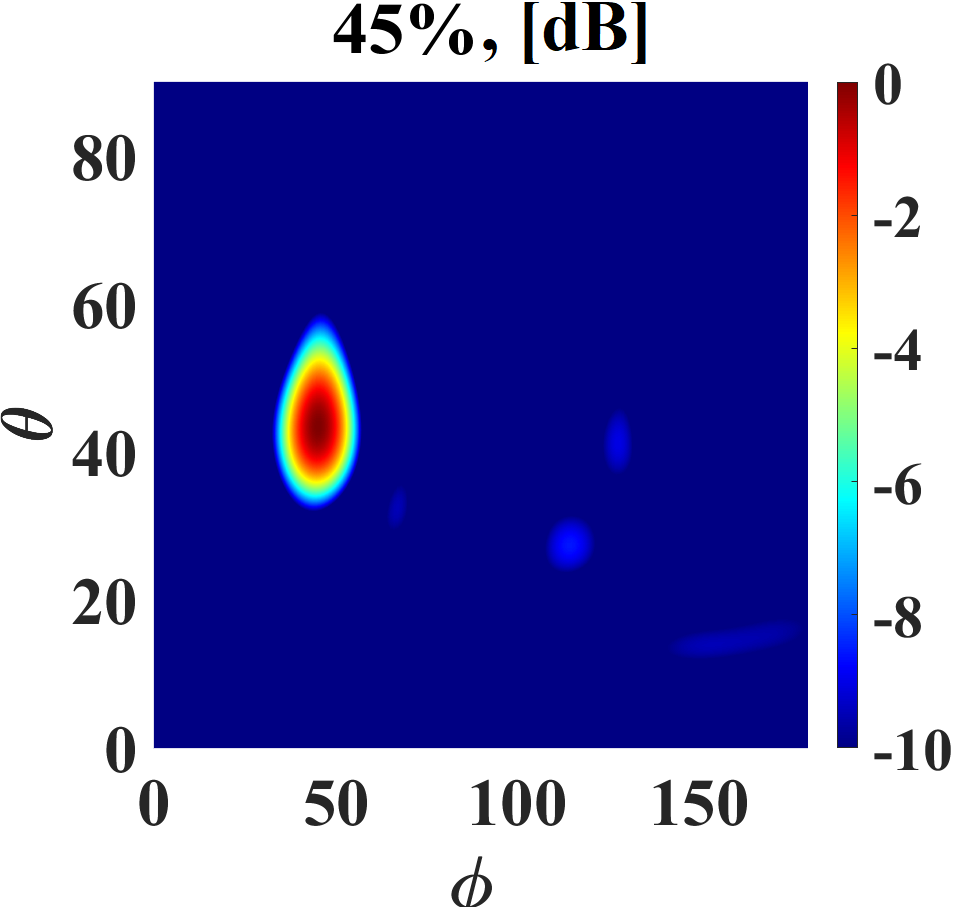}}
	\subfigure[Sketch of the metasurface coding under Independent-Deterministic errors and radiation pattern for three error percentages.\label{ID}]
	{\includegraphics[width=0.5\columnwidth]{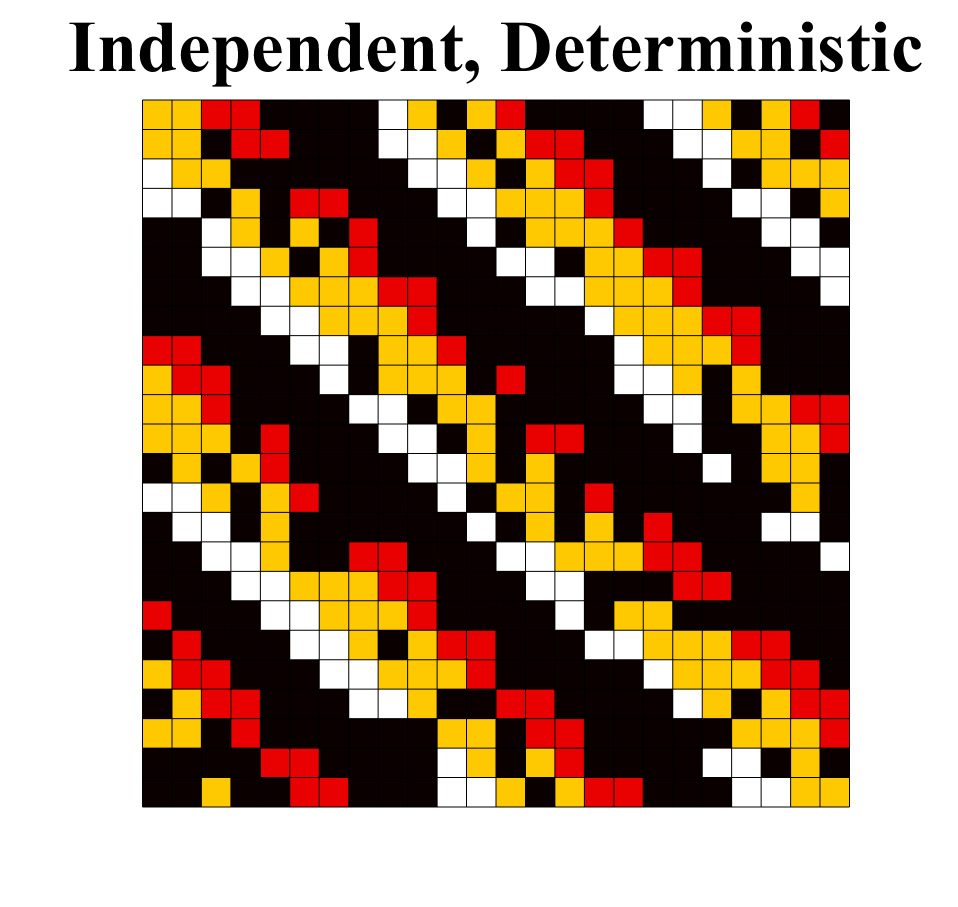}
		\includegraphics[width=0.5\columnwidth]{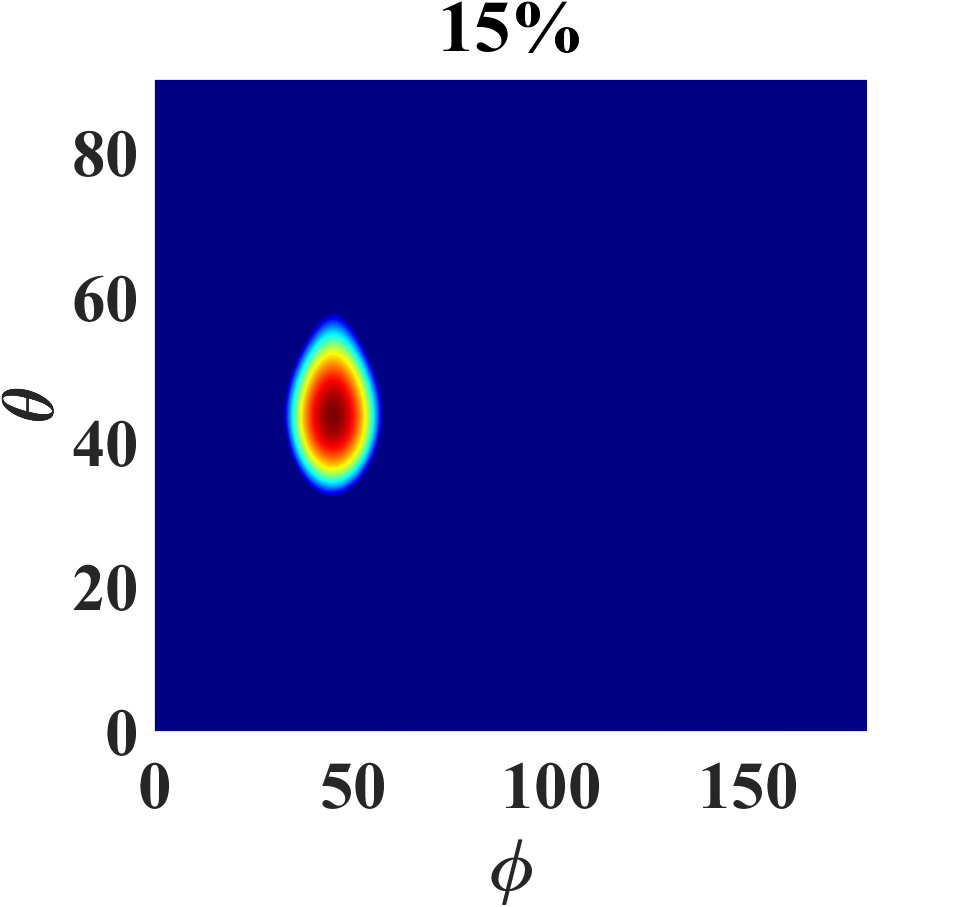}
		\includegraphics[width=0.5\columnwidth]{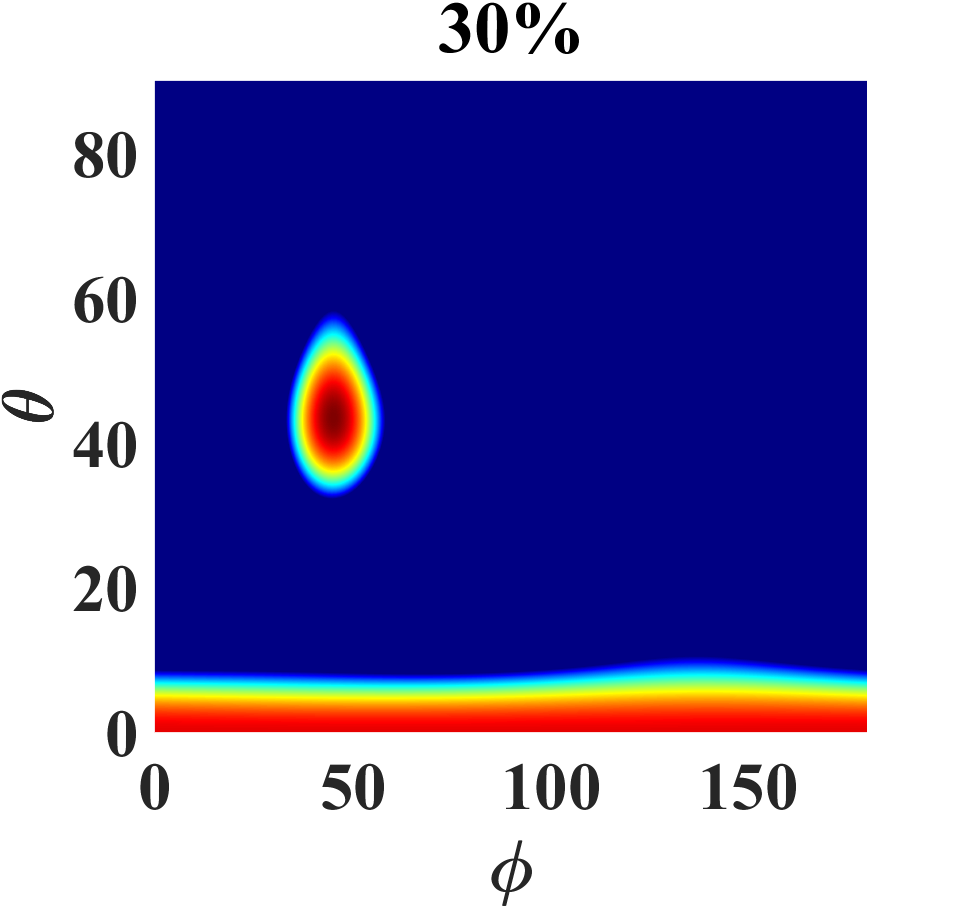}
		\includegraphics[width=0.5\columnwidth]{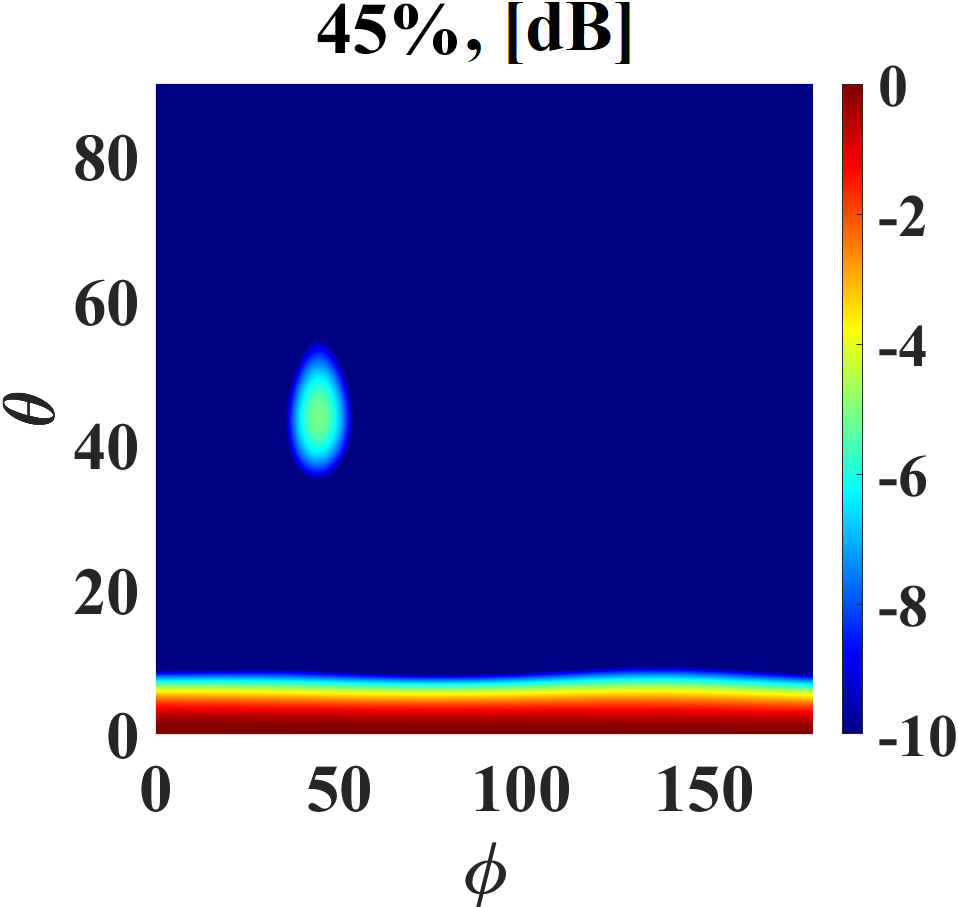}}
	\caption{Qualitative analysis of performance degradation of a beam steering metasurface for different error scenarios.}
				\vspace{-0.4cm}
	\label{mixed}
\end{figure*}

Figure \ref{mixed} demonstrates how different types of error and their spatial distribution can have a significantly different impact. The far field is plotted for increasing fault rates for four representative combinations of error type and spatial distributions. It is observed that the metasurface points most of the energy towards the intended direction of $\theta_r=\phi_r=\pi/4$ for relatively low error percentages and starts losing its functionality as the percentage increases. 

The differences between the distinct types of errors are clearly distinguishable. For instance, Figure \ref{CS} shows the far field for the metasurface with stuck-at errors clustered around the center. We can see that increasing the error ratio gives more power to the side lobes and decreases the width of the main lobe. The clustered-biased scenario shown in Figure \ref{CB}, on the other hand, illustrates that biasing errors are less impactful because, in the end, the phase gradient is largely conserved. Similar studies performed for the (independent, out of state) and (independent, deterministic) scenarios, shown in Figures \ref{IO} and \ref{ID}, respectively, allow to conclude that completely random errors tend to average out and minimize impact, whereas deterministic errors tend to destroy the functionality by increasing the importance of specular reflection, which becomes the main lobe for more than $30\%$ of error ratio.

\hl{Here we set the deterministic error to be $s_0$, which ends up with a strong secondary lobe at $\theta=0$ and arbitrary $\phi$. This reflection angle is independent of the determined value of the error and we would obtain the same results with any other determined value ($s_1,s_2,s_3$). This reflection, namely specular, is only characterized by the incident angle. In other words, deterministic errors react as a mirror reflecting the incident wave according to Snell's law ($\theta_i=\theta_r$ and $\phi_i=\phi_r$) irrespective to the erroneous value, as long as it is the same in all erroneous unit cells. In our case (normal incident of plane wave), where we put $\theta_i=0$ and arbitrary $\phi_i$, we obtain $\theta_r=0$ and arbitrary $\theta_r=0$.}

The next sections complement this qualitative analysis with the evaluation of performance metrics for different combinations of type of error and spatial distribution. For the sake of brevity, we consider the combinations outlined in Table \ref{tab:table1}. The rest of the possible combinations have been evaluated, but are not shown due to their behavioral similarity with the combinations from Table \ref{tab:table1}.

			\vspace{-0.2cm}
\begin{table}[h!]
	\begin{center}
		\caption{Error scenario acronyms.}
			\vspace{-0.2cm}
		\label{tab:table1}
		\begin{tabular}{|c|c|} 
			\hline
			\textbf{Error scenario} & \textbf{acronym} \\
			\hline
			Clustered-Stuck & CS \\
			Clustered-Out of State & CO \\
			Clustered-Deterministic & CD \\
			Clustered-Biased & CB \\
			Independent-Stuck & IS \\
			Independent-Out of State & IO \\
			Independent-Deterministic & ID \\
			Independent-Biased & IB \\
			\hline

		\end{tabular}
		\vspace{-0.7cm}
	\end{center}
\end{table}

\subsection{Directivity}
\begin{figure}[!h]
	\centering
	\includegraphics[width=1\columnwidth]{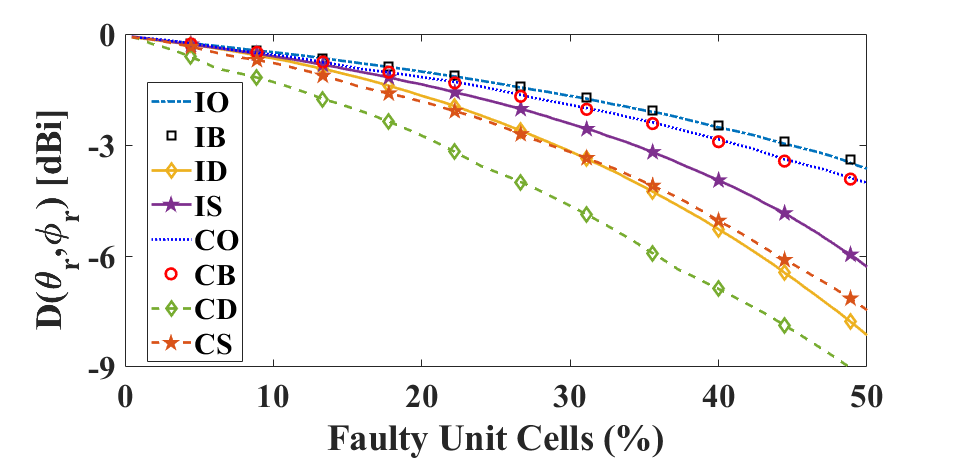}
	\vspace{-0.4cm}
	\caption{Directivity on target direction $D(\theta_r, \phi_r)$ of the beam steering metasurface as a function of the error percentage for different error types and spatial distribution.}
	\label{DT}
	\vspace{-0.15cm}
\end{figure}

Figure \ref{DT} illustrates the impact of the different types of errors by plotting the Directivity at the desired reflection angle $D(\theta_r,\phi_r)$ over the error percentage. As expected, the most detrimental type of error is deterministic because all wrong values are mapped to the same phase, which has a more detrimental effect in the beam steering case due to its phase-gradient requirements. This reasoning also implies that different types of errors may have a completely different impact on metasurfaces implementing different functionalities: for instance, absorbers may set the same value to all unit cells and, therefore, deterministic errors may not reduce performance vitally. Assuming a 3 dB threshold as the acceptable performance degradation, we observe that $CD$ is the most degrading option as error rates beyond 20\% cannot be tolerated, while Out of state and Biased errors ($IO$, $IB$, $CO$ and $CB$) guarantee good performance beyond 40\% of erroneous cells.

\begin{figure}[!h]
	\centering
	\includegraphics[width=1\columnwidth]{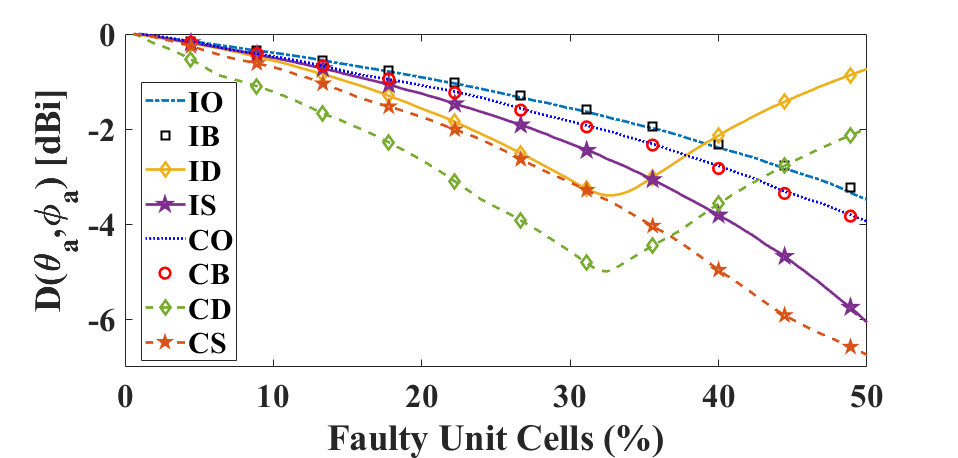}
	\vspace{-0.4cm}
	\caption{Directivity on the direction of maximum reflection $D(\theta_a, \phi_a)$ of the beam steering metasurface as a function of the error percentage for different error types and spatial distribution.}
	\label{DM}
	\vspace{-0.15cm}
\end{figure}

\begin{figure}[!h]
	\centering
	\includegraphics[width=.32\columnwidth]{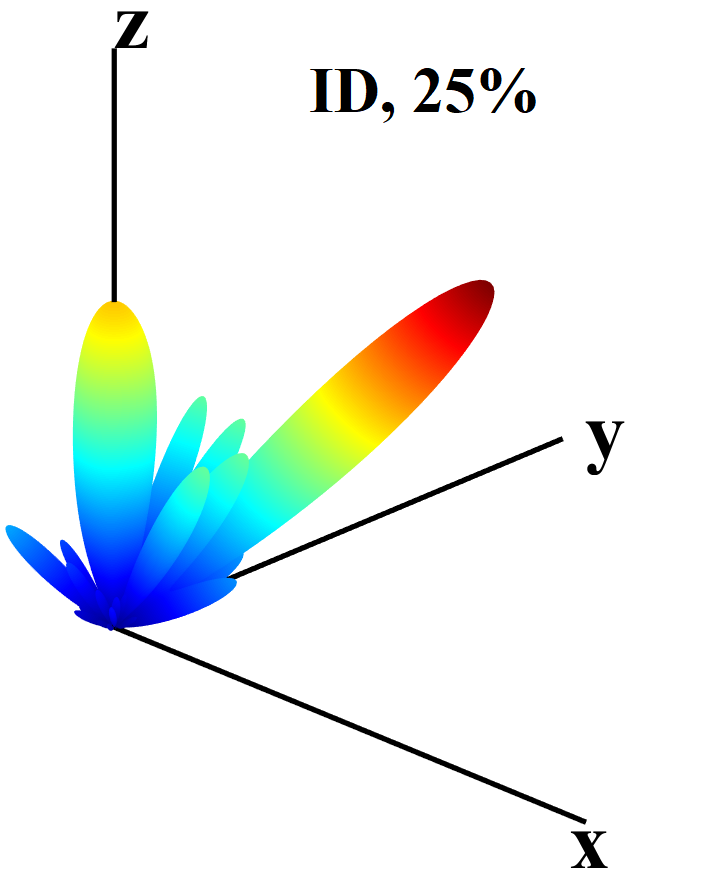}
	\includegraphics[width=.32\columnwidth]{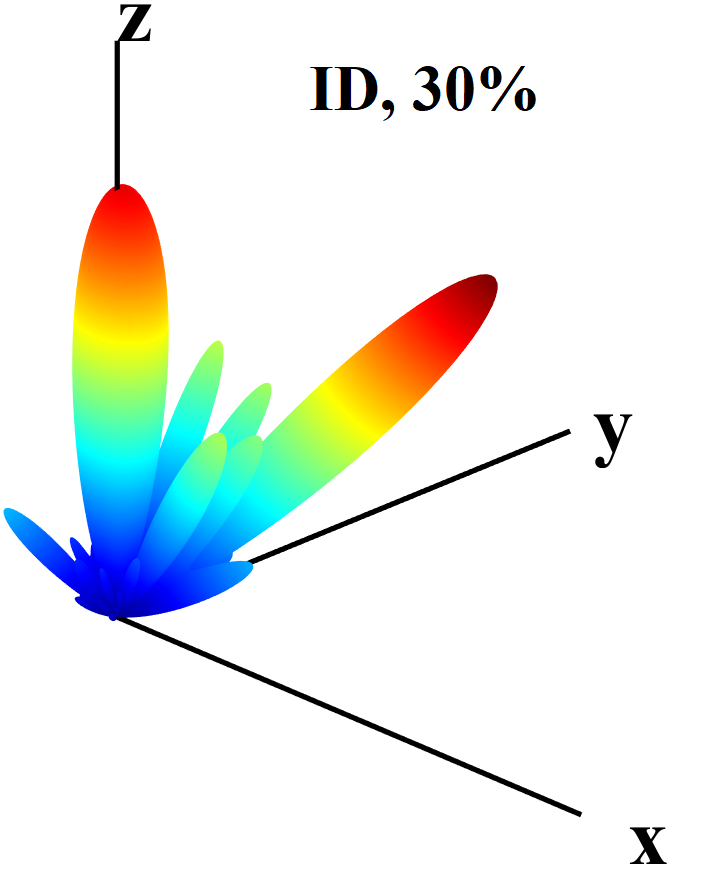}
	\includegraphics[width=.32\columnwidth]{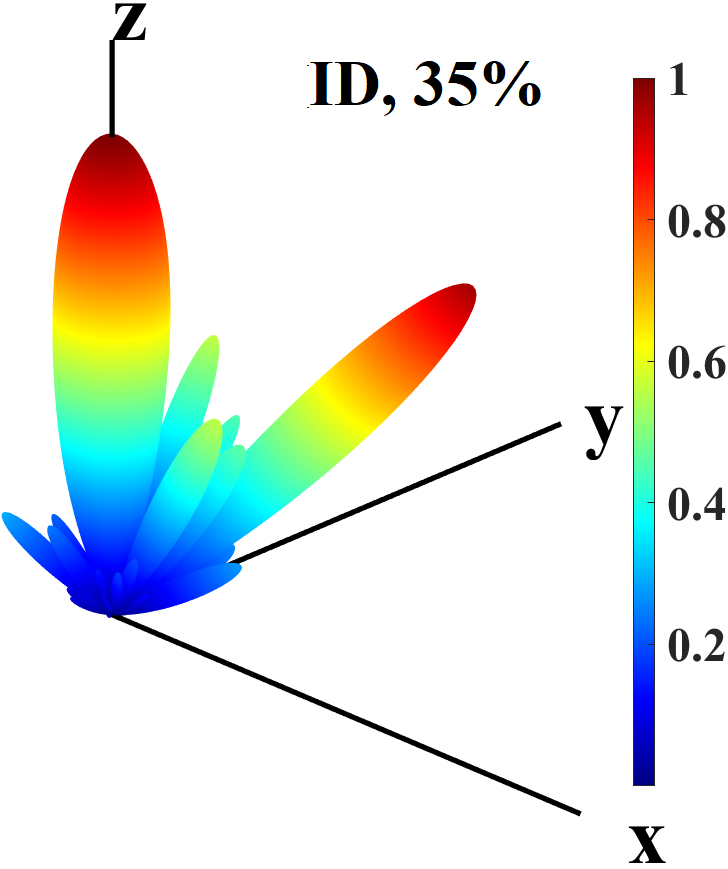}
	\caption{\hl{The normalized radiation pattern for ID errors as an example of an emerging secondary lobe in the wrong direction. The color bar is common to all figures.}}
	\label{com}
	\vspace{-0.15cm}
\end{figure}

The story is different for the directivity at the direction of maximum radiation, $D(\theta_a,\phi_a)$, as shown in Figure \ref{DM}. In this case, the curves of $D(\theta_a,\phi_a)$ corresponding to deterministic errors with fault rates beyond 33\% start to increase regardless of their spatial distribution. The reason for this behavior underlies behind the fact that deterministic errors posses the same reflection phase and, by the accumulation of many of these errors, a secondary beam starts to grow. The 33\% error figure represents the inflection point where the secondary lobe becomes the main lobe. This is illustrated in Figure \ref{com} for the independent-deterministic combination of errors, revealing how the secondary lobe emerges at $\theta=0$ and becomes the main beam.

\subsection{Target Deviation}
\begin{figure}[!h]
	\centering
	\includegraphics[width=1\columnwidth]{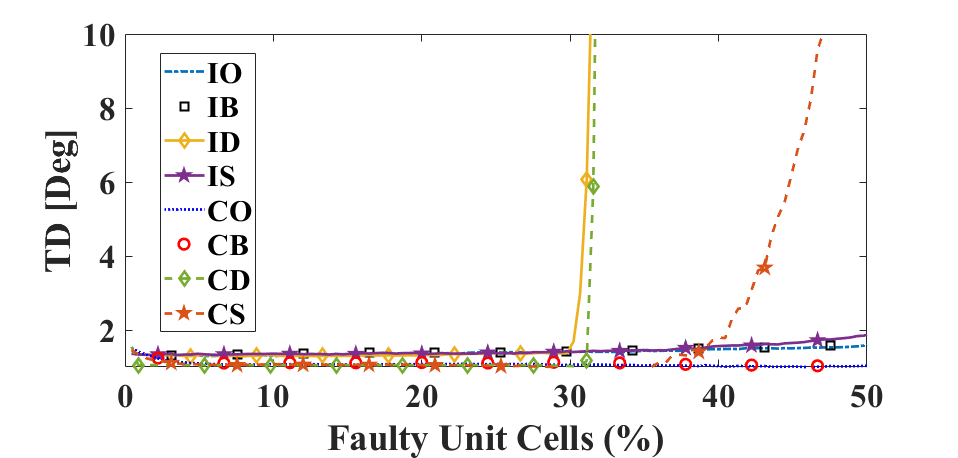}
	\vspace{-0.4cm}
	\caption{Deviation from the target versus the percentage of the faulty unit cells for different error types and spatial distribution.}
	\label{TD}
	\vspace{-0.15cm}
\end{figure}

In this part, we want to emphasize the importance of the differences between the desired and actual position of the main beam. Figure \ref{TD} shows the accuracy of the beam steering metasurface versus different error scenarios. As mentioned above, deterministic errors may make the main beam shift from the target direction to specular reflection. This is the main cause of the jump observed at 33\% for $TD$. Beside $CS$ that forces the main beam to deviate, other kinds of errors are not affecting $TD$ considerably. In fact, $TD$ takes values below 2 degrees, which is acceptable for most applications. This suggests that loss of directivity or increase of side lobe may be more critical in the present scenario. 

\subsection{Half-Power Beamwidth}
\begin{figure}[!h]
	\centering
	\includegraphics[width=1\columnwidth]{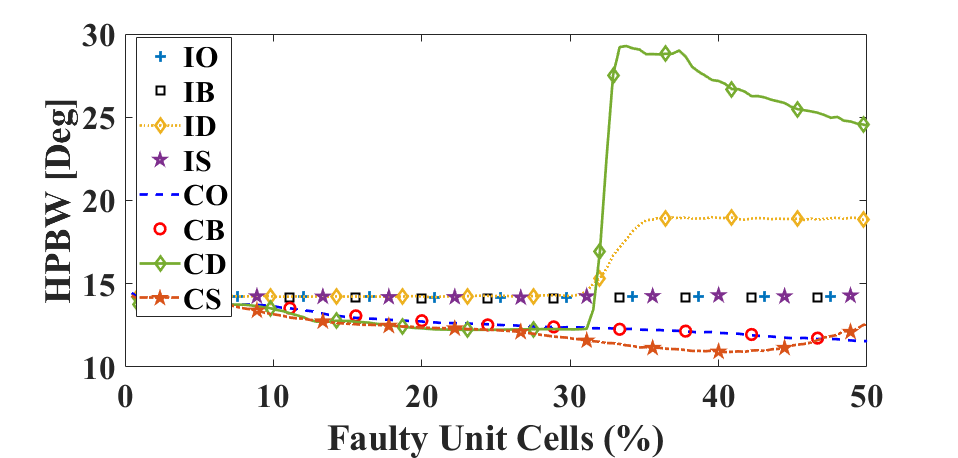}
	\vspace{-0.4cm}
	\caption{Half power beam width versus the percentage of the faulty unit cells for different error types and spatial distribution.}
	\label{BW}
	\vspace{-0.15cm}
\end{figure}

From Figure \ref{BW} it can be inferred that, on one hand, the $HPBW$ is not affected by the $IS$, $IO$ and $IB$ combinations of errors. However, their associated types with cluster distributions, $CS$, $CO$, and $CB$ smoothly decrease the $HPBW$. We can relate this effect to the consolidated spot of errors leading to a united constructive response. On the other hand, for Deterministic cases (i.e., $ID$ and $CD$), \hl{a jump at 33\% is observed due to the deterministic values of errors. At this turning point, anomalous reflection (i.e. the desired behaviour) is very weak, while specular reflection (emerged from deterministic errors) is strong enough to take over as the main beam. This is the reason for the jumps for $HPBW$ in Figure {\ref{BW}}. In other words, before 33\% we have the $HPBW$ of anomalous reflection and after that of specular reflection.}

\subsection{Side-Lobe Level}
\label{sec:Side-Lobe Level}
\begin{figure}[!h]
	\centering
	\includegraphics[width=1\columnwidth]{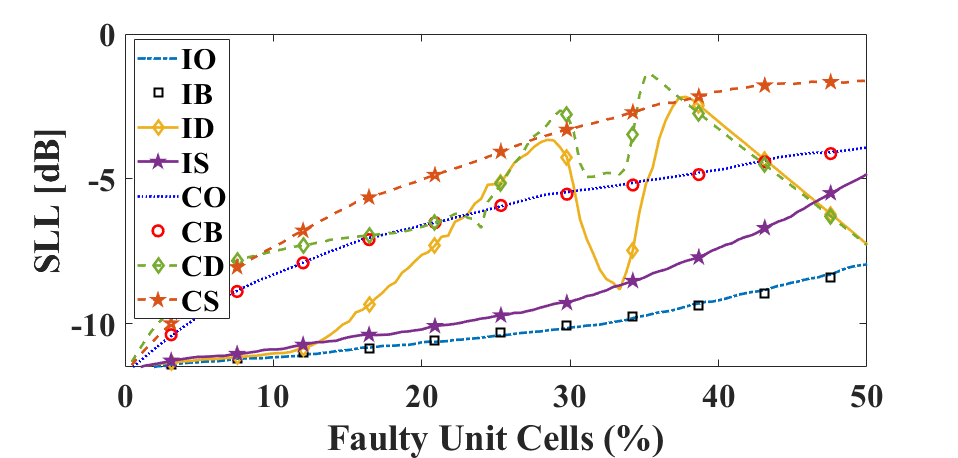}
	\vspace{-0.4cm}
	\caption{Secondary lobe level versus the percentage of the faulty unit cells for different error types and spatial distribution.}
	\label{SL}
	\vspace{-0.15cm}
\end{figure}

Figure \ref{SL} illustrates the performance degradation in terms of side lobe level. It is quickly observed that $SLL$ is the most sensitive performance metric so far. 

Fundamentally, $SLL$ grows proportionally to the ratio of errors for all kinds. Among the monotonically increasing cases, clustered distributions seem to have a higher impact in this metric, whereas independent and out of state errors are the least relevant. Since random coding \cite{Moccia2017} leads to random scattering, we argue that uncorrelated random errors generate scattering that does not accumulate as a large secondary lobe. On the other hand, clustered and deterministic errors, which tend to group unit cells together and to apply a uniform state, lead to large secondary lobes. 

\hl{Deterministic errors lead to a characteristic behaviour that we elucidated with several simulations at specific points for $CD$ and $ID$. The dip appearing between 29\% and 37\% is due to the fact that a secondary lobe starts to disappear while another minor lobe begins to rise. At 32\%, this minor lobe takes the place of secondary lobe then at 37\% main and secondary beams are exchanging their roles and $SLL$ starts to decline once again.}

\subsection{Side-Lobe Accumulated Energy}
\begin{figure}[!h]
	\centering
	\includegraphics[width=1\columnwidth]{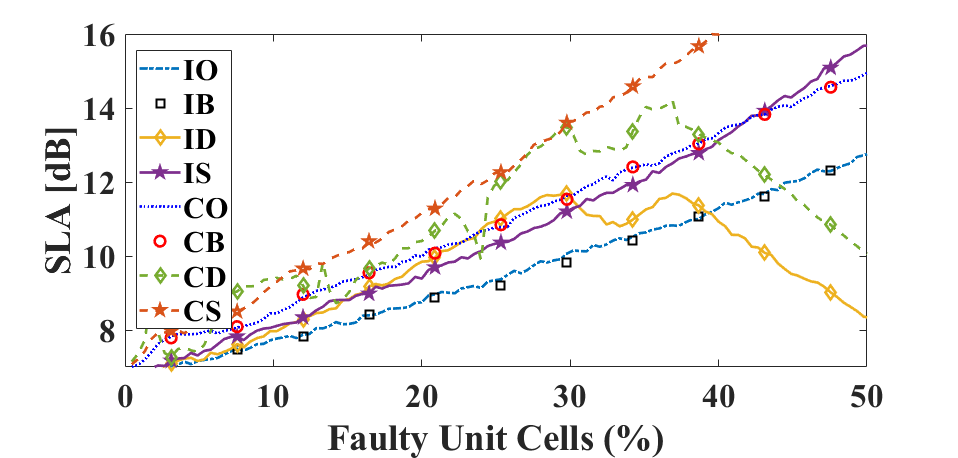}
	\vspace{-0.4cm}
	\caption{Accumulated energy of the side lobes versus percentage of the faulty unit cells for different error types and spatial distribution.}
	\label{OL}
	\vspace{-0.15cm}
\end{figure}

$SLL$ cannot accurately describe the distribution of energy around the main lobe, so we evaluate the $SLA$ metric in Figure \ref{OL} to consider the impact of all the minor lobes. This figure demonstrates that, as expected, increasing fault rates reduce the power at the main beam and distribute it around a set of side-lobes. The main lobe is debilitated very similarly for all error combinations. \hl{Likewise, deterministic errors exhibit abnormal behaviour for $SLA$ showing a dip between [29\%-32\%] and decline after 37\%. The reason being of such trend is similar to that discussed in Section {\ref{sec:Side-Lobe Level}}}.

\section{Conclusion}
\label{sec:conclusions}
We have proposed an error model and a methodology for error analysis in metasurfaces. Beam steering metasurfaces, where the functionality depends on the phase gradient, are robust against spatially uncorrelated errors with random values and to attacks that only bias the state of the unit cells. On the contrary, clustered errors that set all the unit cells to the same state are very detrimental. These results show the value of the error analysis and suggest that the error model is comprehensive enough to cover all possible cases. Future works will further analyze the impact of errors in metasurfaces with different sizes or functionalities, to then derive useful design guidelines and power-gating directives.

\section*{Acknowledgment}
This work has been supported by the European Commission under grant H2020-FETOPEN-736876 (VISORSURF) and by ICREA under the ICREA Academia programme.


\begin{IEEEbiography}[{\includegraphics[width=1in,height=1.25in,clip,keepaspectratio]{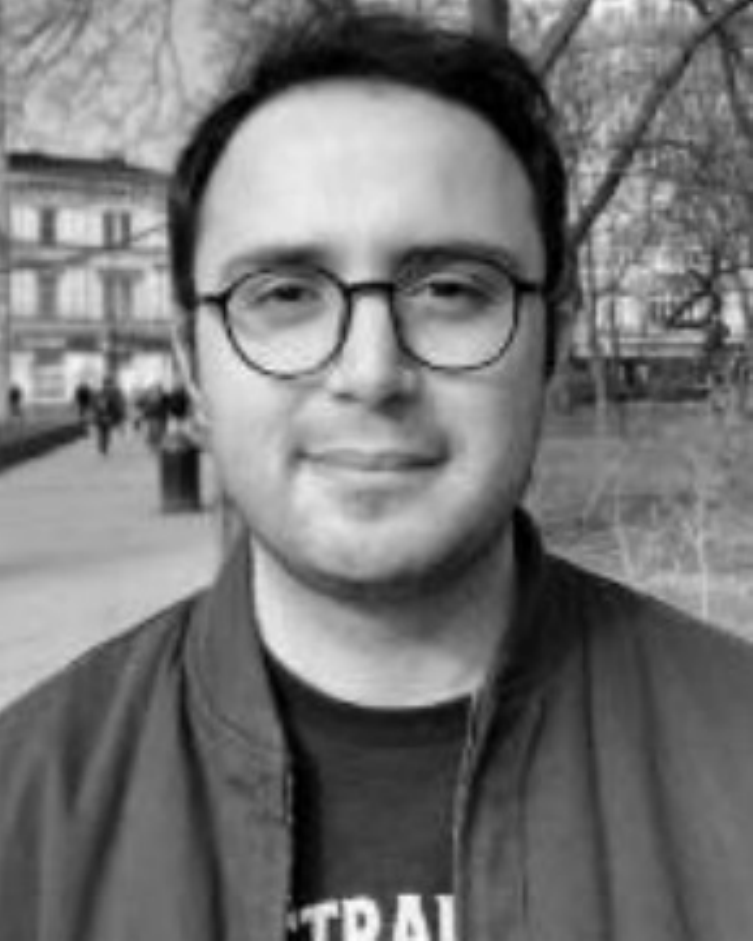}}]{Hamidreza Taghvaee}
	received the M.Sc. in Telecommunication Engineering majored in Field$\&$Wave from K. N. Toosi University of Technology, Tehran, Iran, in 2016. He was an optical engineer in transmission department of Huawei leading field service engineers for network development. In 2018 he started a PhD degree at Universitat Polit\`{e}cnica de Catalunya (UPC), where he joined the NaNoNetworking Center in Catalunya being researcher in the VISORSURF FET-OPEN project. Currently, he is a visiting researcher at Aalto University where he conducts multidisciplinary research in programmable metasurfaces. His main research interests includes Electromagnetic, Metamaterial and Antenna. 
\end{IEEEbiography}

\begin{IEEEbiography}[{\includegraphics[width=1in,height=1.25in,clip,keepaspectratio]{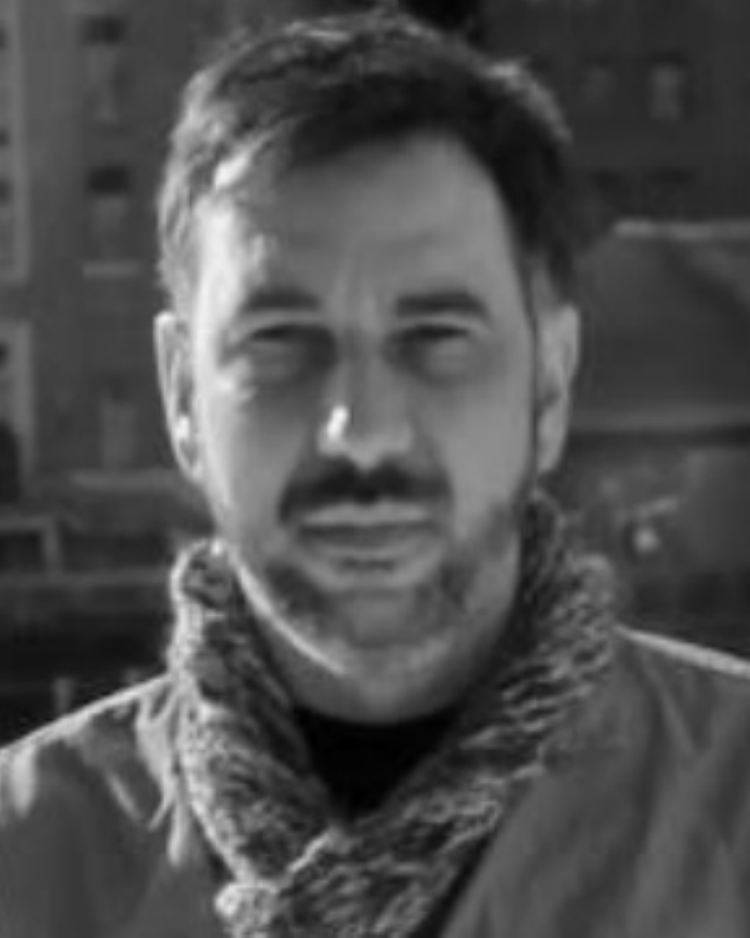}}]{Albert Cabellos-Aparicio}
	received a MSc (2005) and PhD (2008) degree in Computer Science Engineering from the Technical University of Catalonia (www.upc.edu). In september 2005 he became an assistant professor of the Computer Architecture Department and as a researcher in the Broadband Communications Group (http://cba.upc.edu/). In 2010 he joined the NaNoNetworking Center in Catalunya (http://www.n3cat.upc.edu) where he is the Scientific Director. He is an editor of the Elsevier Journal on Nano Computer Network and founder of the ACM NANOCOM conference, the IEEE MONACOM workshop and the N3Summit. He has also founded the LISPmob open-source initiative (http://lispmob.org) along with Cisco. He has been a visiting researcher at Cisco Systems and Agilent Technologies and a visiting professor at the Royal Institute of Technology (KTH) and the Massachusetts Institute of Technology (MIT). He has given more than 10 invited talks (MIT, Cisco, INTEL, MIET, Northeastern Univ. etc.) and co-authored more than 40 journal and 100 conference papers. His main research interests are future architectures for the Internet and nano-scale communications.
	
\end{IEEEbiography}

\begin{IEEEbiography}[{\includegraphics[width=1in,height=1.25in,clip,keepaspectratio]{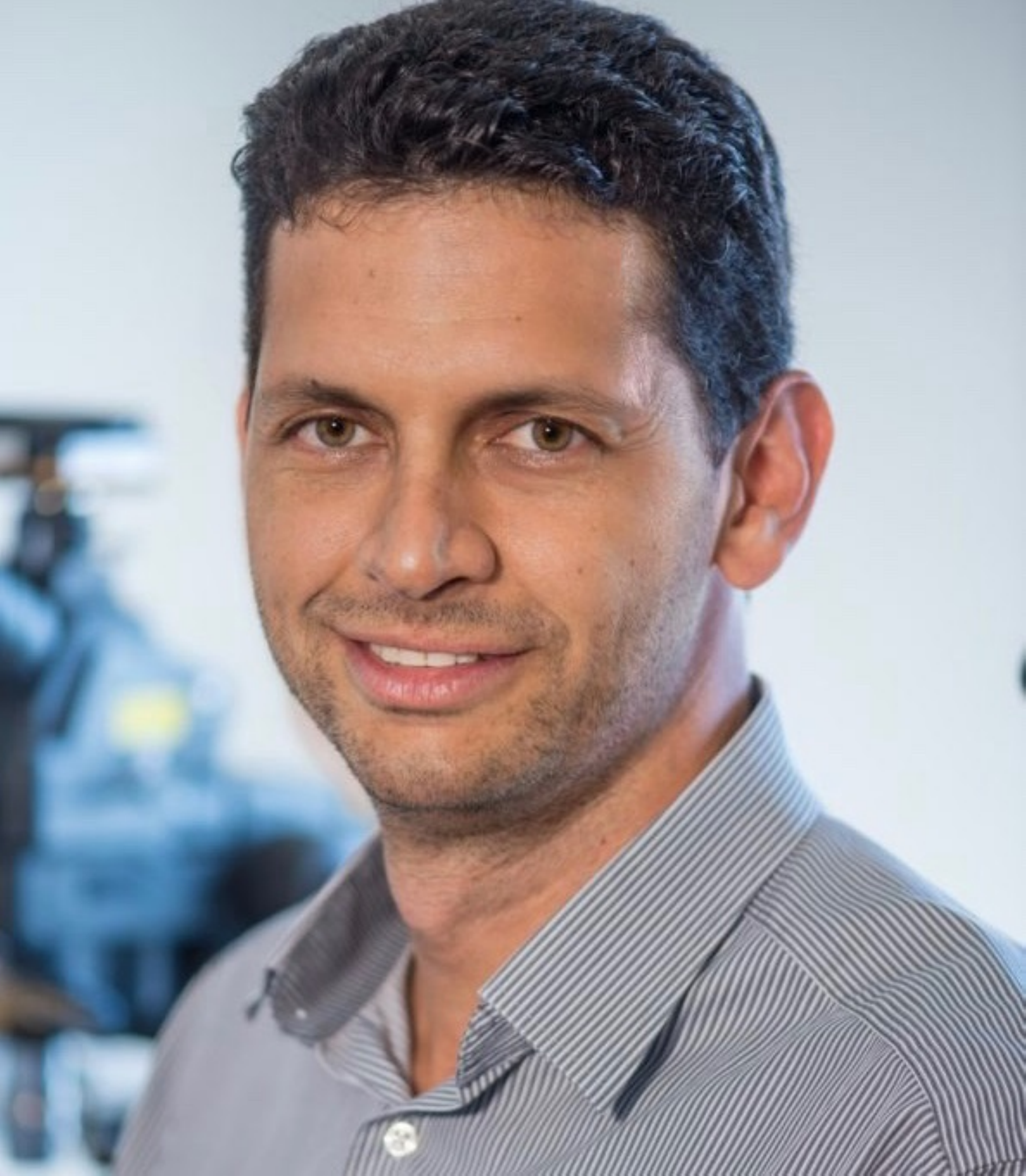}}]{Julius Georgiou}
	(IEEE M’98-SM’08) is an Associate Professor at the University of Cyprus. He received his M.Eng degree in Electrical and Electronic Engineering and Ph.D. degree from Imperial College London in 1998 and 2003, respectively. For two years he worked as Head of Micropower Design in a technology start-up company, Toumaz Technology. In 2004, he joined the Johns Hopkins University as a Postdoctoral Fellow, before becoming a faculty member at the University of Cyprus in 2005, to date. Prof. Georgiou is a member of the IEEE Circuits and Systems Society, is the Chair of the IEEE Biomedical and Life Science Circuits and Systems (BioCAS) Technical Committee, as well as a member of the IEEE Circuits and Systems Society Analog Signal Processing Technical Committee. He served as the General Chair of the 2010 IEEE Biomedical Circuits and Systems Conference and is the Action Chair of the EU COST Action ICT-1401 on “Memristors-Devices, Models, Circuits, Systems and Applications - MemoCIS”. Prof. Georgiou has been selected as an IEEE Circuits and Systems Society Distinguished Lecturer for 2016-2017. He is also an Associate Editor of the IEEE Transactions on Biomedical Circuits and Systems and Associate Editor of the Frontiers in Neuromorphic Engineering Journal. He is a recipient of a best paper award at the IEEE ISCAS 2011 International Symposium and a best Paper awarded at the IEEE BioDevices 2008 Conference. In 2016, he received the 2015 ONE Award from the President of the Republic of Cyprus for his research accomplishments related to the development of an “Infrared Fluorescence-Based Cancer Screening Capsule for the Small Intestine”. His research interests include Low-power analog and digital ASICs, implantable biomedical devices, bioinspired electronic systems, electronics for space, brain-computer-interfaces (BCIs), memristive devices, electronics for adaptive metamaterials, inertial and optical sensors and related systems.
\end{IEEEbiography}

\begin{IEEEbiography}
	[{\includegraphics[width=1in,height=1.25in,clip,keepaspectratio]{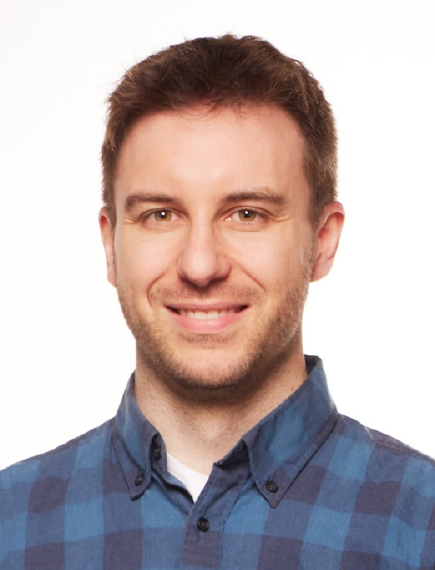}}]{Sergi Abadal}
	received the PhD in Computer Architecture from the Department of Computer Architecture, Universitat Polit\`{e}cnica de Catalunya (UPC), Barcelona, Spain, in July 2016. Previously, he had obtained the M.Sc. in Information and Communication Technologies (MINT) and the B.Sc. in Telecommunication Engineering from the Telecommunication Engineering School (Telecom BCN), UPC, Barcelona, Spain, in 2011 and 2010, respectively. He has held several visiting positions in Georgia Tech in 2009, University of Illinois Urbana-Champaign in 2015 and 2016, and the Foundation of Research and Technology – Hellas in 2018. Currently, he is coordinator of the WIPLASH FET-OPEN project and Area Editor of the Nano Communication Networks (Elsevier) Journal, where he was named Editor of the Year 2019. Abadal has also served as TPC member of more than 10 conferences and organized 3 special sessions, and published over 60 articles in top-tier journals and conferences. Abadal was the recipient of the Nano Communication Networks Young Researcher Award in 2019, INTEL Doctoral Fellowship in 2013, Accenture Award for M.Sc. students in 2012 and Vodafone Fellowship for outgoing B.Sc students in 2010. His current research interests are in the areas of chip-scale wireless communications, including channel modeling and protocol design, the application of these techniques for the creation next-generation wireless networks for on-chip or within metamaterial networks, and their implications at the architecture/system design level. 
\end{IEEEbiography}


\begin{thebibliography}{10}
	\providecommand{\url}[1]{#1}
	\csname url@samestyle\endcsname
	\providecommand{\newblock}{\relax}
	\providecommand{\bibinfo}[2]{#2}
	\providecommand{\BIBentrySTDinterwordspacing}{\spaceskip=0pt\relax}
	\providecommand{\BIBentryALTinterwordstretchfactor}{4}
	\providecommand{\BIBentryALTinterwordspacing}{\spaceskip=\fontdimen2\font plus
		\BIBentryALTinterwordstretchfactor\fontdimen3\font minus
		\fontdimen4\font\relax}
	\providecommand{\BIBforeignlanguage}[2]{{%
			\expandafter\ifx\csname l@#1\endcsname\relax
			\typeout{** WARNING: IEEEtran.bst: No hyphenation pattern has been}%
			\typeout{** loaded for the language `#1'. Using the pattern for}%
			\typeout{** the default language instead.}%
			\else
			\language=\csname l@#1\endcsname
			\fi
			#2}}
	\providecommand{\BIBdecl}{\relax}
	\BIBdecl
	
	\bibitem{Engheta2006}
	N.~Engheta and R.~W. Ziolkowski, \emph{{Metamaterials: physics and engineering
			applications}}, J.~W. Sons, Ed., 2006.
	
	\bibitem{Glybovski2016}
	S.~B. Glybovski, S.~A. Tretyakov, P.~A. Belov, Y.~S. Kivshar, and C.~R.
	Simovski, ``{Metasurfaces: From microwaves to visible},'' \emph{Physics
		Reports}, vol. 634, pp. 1--72, 2016.
	
	\bibitem{Taghvaee2014}
	H.~Taghvaee, M.~Seyyedi, and A.~Rezaee, ``{Design of a Metamaterial Dual Band
		Absorber},'' in \emph{The third Iranian Conference on Engineering
		Electromagnetic ICEEM}, vol.~3.\hskip 1em plus 0.5em minus 0.4em\relax
	Tehran: Iranian Scientific Society of Engineering Electromagnetics, 2014.
	
	\bibitem{Chen2016}
	H.~T. Chen, A.~J. Taylor, and N.~Yu, ``{A review of metasurfaces: Physics and
		applications},'' \emph{Reports on Progress in Physics}, vol.~79, no.~7, 2016.
	
	\bibitem{Vellucci2017}
	S.~Vellucci, A.~Monti, M.~Barbuto, A.~Toscano, and F.~Bilotti, ``{Satellite
		Applications of Electromagnetic Cloaking},'' \emph{IEEE Transactions on
		Antennas and Propagation}, vol.~65, no.~9, pp. 4931--4934, 2017.
	
	\bibitem{Huang2017}
	C.~Huang, B.~Sun, W.~Pan, J.~Cui, X.~Wu, and X.~Luo, ``{Dynamical beam
		manipulation based on 2-bit digitally-controlled coding metasurface},''
	\emph{Scientific Reports}, vol.~7, no. January, pp. 1--8, 2017.
	
	\bibitem{Li2017b}
	L.~Li, T.-J. Cui, W.~Ji, S.~Liu, J.~Ding, X.~Wan, Y.~{Bo Li}, M.~Jiang, C.~W.
	Qiu, and S.~Zhang, ``{Electromagnetic reprogrammable coding-metasurface
		holograms},'' \emph{Nature Communications}, vol.~8, no.~1, pp. 1--7, 2017.
	
	\bibitem{Tcvetkova2018}
	S.~N. Tcvetkova, D.-H. Kwon, A.~D{\'{i}}az-Rubio, and S.~A. Tretyakov,
	``{Near-perfect conversion of a propagating plane wave into a surface wave
		using metasurfaces},'' \emph{Physical Review B}, vol.~97, p. 115447, 2018.
	
	\bibitem{Qu2015}
	S.~W. Qu, W.~W. Wu, B.~J. Chen, H.~Yi, X.~Bai, K.~B. Ng, and C.~H. Chan,
	``{Controlling Dispersion Characteristics of Terahertz Metasurface},''
	\emph{Scientific Reports}, vol.~5, p. 9367, 2015.
	
	\bibitem{Hosseininejad2019}
	S.~E. Hosseininejad, K.~Rouhi, M.~Neshat, R.~Faraji-Dana, A.~Cabellos-Aparicio,
	S.~Abadal, and E.~Alarc{\'{o}}n, ``{Reprogrammable Graphene-based Metasurface
		Mirror with Adaptive Focal Point for THz Imaging},'' \emph{Scientific
		Reports}, vol.~9, p. 2868, 2019.
	
	\bibitem{Liu2016a}
	S.~Liu, T.~J. Cui, Q.~Xu, D.~Bao, L.~Du, X.~Wan, W.~X. Tang, C.~Ouyang, X.~Y.
	Zhou, H.~Yuan, H.~F. Ma, W.~X. Jiang, J.~Han, W.~Zhang, and Q.~Cheng,
	``{Anisotropic coding metamaterials and their powerful manipulation of
		differently polarized terahertz waves},'' \emph{Light: Science and
		Applications}, vol.~5, no.~5, pp. e16\,076----11, 2016.
	
	\bibitem{Taghvaee2017}
	H.~R. Taghvaee, F.~Zarrinkhat, and M.~S. Abrishamian, ``{Terahertz Kerr
		nonlinearity analysis of a microribbon graphene array using the harmonic
		balance method},'' \emph{Journal of Physics D: Applied Physics}, vol.~50,
	no.~25, 2017.
	
	\bibitem{Qu2017}
	S.~W. Qu, H.~Yi, B.~J. Chen, K.~B. Ng, and C.~H. Chan, ``{Terahertz Reflecting
		and Transmitting Metasurfaces},'' \emph{Proceedings of the IEEE}, vol. 105,
	no.~6, pp. 1166--1184, 2017.
	
	\bibitem{ChenXZ2012}
	X.~Z. Chen, L.~L. Huang, H.~M{\"{u}}hlenbernd, G.~X. Li, and B.~F. Bai,
	``{Dual-polarity plasmonic metalens for visible light},'' \emph{Nature
		Communications}, no.~3, p. 1198, 2012.
	
	\bibitem{Li2015}
	Z.~Li, K.~Yao, F.~Xia, S.~Shen, J.~Tian, and Y.~Liu, ``{Graphene Plasmonic
		Metasurfaces to Steer Infrared Light},'' \emph{Scientific Reports}, vol.~5,
	pp. 1--9, 2015.
	
	\bibitem{Oliveri2015}
	G.~Oliveri, D.~Werner, and A.~Massa, ``{Reconfigurable electromagnetics through
		metamaterials - A Review},'' \emph{Proceedings of the IEEE}, vol. 103, no.~7,
	pp. 1034--1056, 2015.
	
	\bibitem{Cui2014}
	T.~J. Cui, M.~Q. Qi, X.~Wan, J.~Zhao, Q.~Cheng, K.~T. Lee, J.~Y. Lee, S.~Seo,
	L.~J. Guo, Z.~Zhang, Z.~You, and D.~Chu, ``{Coding metamaterials, digital
		metamaterials and programmable metamaterials},'' \emph{Light: Science and
		Applications}, vol.~3, no.~10, pp. 1--9, 2014.
	
	\bibitem{Wan2016}
	X.~Wan, M.~Q. Qi, T.~Y. Chen, and T.~J. Cui, ``{Field-programmable beam
		reconfiguring based on digitally-controlled coding metasurface},''
	\emph{Scientific Reports}, vol.~6, p. 20663, 2016.
	
	\bibitem{AbadalACCESS}
	S.~Abadal, C.~Liaskos, A.~Tsioliaridou, S.~Ioannidis, A.~Pitsillides,
	J.~Sol{\'{e}}-Pareta, E.~Alarc{\'{o}}n, and A.~Cabellos-Aparicio,
	``{Computing and Communications for the Software-Defined Metamaterial
		Paradigm: A Context Analysis},'' \emph{IEEE Access}, vol.~5, pp. 6225--6235,
	2017.
	
	\bibitem{Tasolamprou8788546}
	A.~C. {Tasolamprou}, A.~{Pitilakis}, S.~{Abadal}, O.~{Tsilipakos},
	X.~{Timoneda}, H.~{Taghvaee}, M.~{Sajjad Mirmoosa}, F.~{Liu}, C.~{Liaskos},
	A.~{Tsioliaridou}, S.~{Ioannidis}, N.~V. {Kantartzis}, D.~{Manessis},
	J.~{Georgiou}, A.~{Cabellos-Aparicio}, E.~{Alarcón}, A.~{Pitsillides}, I.~F.
	{Akyildiz}, S.~A. {Tretyakov}, E.~N. {Economou}, M.~{Kafesaki}, and C.~M.
	{Soukoulis}, ``Exploration of intercell wireless millimeter-wave
	communication in the landscape of intelligent metasurfaces,'' \emph{IEEE
		Access}, vol.~7, pp. 122\,931--122\,948, 2019.
	
	\bibitem{Liaskos2018a}
	C.~Liaskos, A.~Tsioliaridou, A.~Pitsillides, S.~Ioannidis, and I.~Akyildiz,
	``{Using any Surface to Realize a New Paradigm for Wireless
		Communications},'' \emph{Communications of the ACM}, vol.~61, no.~11, pp.
	30--33, 2018.
	
	\bibitem{PhysRevApplied.11.044024}
	F.~Liu, O.~Tsilipakos, A.~Pitilakis, A.~C. Tasolamprou, M.~S. Mirmoosa, N.~V.
	Kantartzis, D.-H. Kwon, J.~Georgiou, K.~Kossifos, M.~A. Antoniades,
	M.~Kafesaki, C.~M. Soukoulis, and S.~A. Tretyakov, ``Intelligent metasurfaces
	with continuously tunable local surface impedance for multiple reconfigurable
	functions,'' \emph{Phys. Rev. Applied}, vol.~11, p. 044024, Apr 2019.
	
	\bibitem{Liaskos2015}
	C.~Liaskos, A.~Tsioliaridou, A.~Pitsillides, I.~F. Akyildiz, N.~V. Kantartzis,
	A.~X. Lalas, X.~Dimitropoulos, S.~Ioannidis, M.~Kafesaki, and C.~M.
	Soukoulis, ``{Design and Development of Software Defined Metamaterials for
		Nanonetworks},'' \emph{IEEE Circuits and Systems Magazine}, vol.~15, no.~4,
	pp. 12--25, 2015.
	
	\bibitem{di2019smart}
	M.~Di~Renzo, M.~Debbah, D.-T. Phan-Huy, A.~Zappone, M.-S. Alouini, C.~Yuen,
	V.~Sciancalepore, G.~C. Alexandropoulos, J.~Hoydis, H.~Gacanin \emph{et~al.},
	``Smart radio environments empowered by reconfigurable ai meta-surfaces: an
	idea whose time has come,'' \emph{EURASIP Journal on Wireless Communications
		and Networking}, vol. 2019, no. 129, 2019.
	
	\bibitem{huang2019reconfigurable}
	C.~Huang, A.~Zappone, G.~C. Alexandropoulos, M.~Debbah, and C.~Yuen,
	``Reconfigurable intelligent surfaces for energy efficiency in wireless
	communication,'' \emph{IEEE Transactions on Wireless Communications},
	vol.~18, no.~8, pp. 4157--4170, 2019.
	
	\bibitem{ma2019smart}
	Q.~Ma, G.~D. Bai, H.~B. Jing, C.~Yang, L.~Li, and T.~J. Cui, ``Smart
	metasurface with self-adaptively reprogrammable functions,'' \emph{Light:
		Science \& Applications}, vol.~8, no.~1, pp. 1--12, 2019.
	
	\bibitem{nwz148}
	H.~C. Zhang, T.~J. Cui, Y.~Luo, J.~Zhang, J.~Xu, P.~H. He, and L.~P. Zhang,
	``Active digital spoof plasmonics,'' \emph{National Science Review}, 10 2019.
	
	\bibitem{HsfNetworkTNET.2019}
	C.~Liaskos, A.~Tsioliaridou, S.~Nie, A.~Pitsillides, S.~Ioannidis, and I.~F.
	Akyildiz, ``On the network-layer modeling and configuration of programmable
	wireless environments,'' \emph{IEEE/ACM Transactions on Networking}, vol.~27,
	no.~4, pp. 1696--1713, 2019.
	
	\bibitem{Liaskos2018b}
	C.~Liaskos, S.~Nie, A.~Tsioliaridou, A.~Pitsillides, S.~Ioannidis, and
	I.~Akyildiz, ``{A New Wireless Communication Paradigm through
		Software-Controlled Metasurfaces},'' \emph{IEEE Communications Magazine},
	vol.~56, no.~9, pp. 162--169, 2018.
	
	\bibitem{perovic2019channel}
	N.~S. Perovi{\'c}, M.~Di~Renzo, and M.~F. Flanagan, ``Channel capacity
	optimization using reconfigurable intelligent surfaces in indoor mmwave
	environments,'' \emph{arXiv preprint arXiv:1910.14310}, 2019.
	
	\bibitem{bjornson2019intelligent}
	E.~Bj{\"o}rnson, {\"O}.~{\"O}zdogan, and E.~G. Larsson, ``Intelligent
	reflecting surface vs. decode-and-forward: How large surfaces are needed to
	beat relaying?'' \emph{arXiv preprint arXiv:1906.03949}, 2019.
	
	\bibitem{Basar2019}
	\BIBentryALTinterwordspacing
	E.~Basar, ``{Large Intelligent Surface-Based Index Modulation: A New Beyond
		MIMO Paradigm for 6G},'' pp. 1--10, 2019. [Online]. Available:
	\url{http://arxiv.org/abs/1904.06704}
	\BIBentrySTDinterwordspacing
	
	\bibitem{badiu2019communication}
	M.-A. Badiu and J.~P. Coon, ``Communication through a large reflecting surface
	with phase errors,'' 2019.
	
	\bibitem{rondinelli1966effects}
	L.~Rondinelli, ``Effects of random errors on the performance of antenna arrays
	of many elements,'' in \emph{1958 IRE International Convention Record},
	vol.~7.\hskip 1em plus 0.5em minus 0.4em\relax IEEE, 1966, pp. 174--189.
	
	\bibitem{zahm1972effects}
	C.~Zahm, ``Effects of errors in the direction of incidence on the performance
	of an adaptive array,'' \emph{Proceedings of the IEEE}, vol.~60, no.~8, pp.
	1008--1009, 1972.
	
	\bibitem{wang1992performance}
	H.~Wang, ``Performance of phased-array antennas with mechanical errors,''
	\emph{IEEE Transactions on Aerospace and Electronic Systems}, vol.~28, no.~2,
	pp. 535--545, 1992.
	
	\bibitem{poli2015dealing}
	L.~Poli, P.~Rocca, N.~Anselmi, and A.~Massa, ``Dealing with uncertainties on
	phase weighting of linear antenna arrays by means of interval-based tolerance
	analysis,'' \emph{IEEE Transactions on Antennas and Propagation}, vol.~63,
	no.~7, pp. 3229--3234, 2015.
	
	\bibitem{8702080}
	H.~{Taghvaee}, S.~{Abadal}, J.~{Georgiou}, A.~{Cabellos-Aparicio}, and
	E.~{Alarcón}, ``Fault tolerance in programmable metasurfaces: The beam
	steering case,'' in \emph{2019 IEEE International Symposium on Circuits and
		Systems (ISCAS)}, May 2019, pp. 1--5.
	
	\bibitem{Li2018}
	A.~Li, S.~Singh, and D.~Sievenpiper, ``{Metasurfaces and their applications},''
	\emph{Nanophotonics}, vol.~7, no.~6, pp. 989--1011, 2018.
	
	\bibitem{Tsilipakos2018a}
	O.~Tsilipakos, A.~C. Tasolamprou, T.~Koschny, M.~Kafesaki, E.~N. Economou, and
	C.~M. Soukoulis, ``{Pairing Toroidal and Magnetic Dipole Resonances in
		Elliptic Dielectric Rod Metasurfaces for Reconfigurable Wavefront
		Manipulation in Reflection},'' \emph{Advanced Optical Materials}, vol.~6,
	no.~22, 2018.
	
	\bibitem{Zhang2017}
	M.~Zhang, W.~Zhang, A.~Q. Liu, F.~C. Li, and C.~F. Lan, ``{Tunable Polarization
		Conversion and Rotation based on a Reconfigurable Metasurface},''
	\emph{Scientific Reports}, vol.~7, no.~1, pp. 1--7, 2017.
	
	\bibitem{Tsilipakos2018}
	O.~Tsilipakos, F.~Liu, A.~Pitilakis, A.~C. Tasolamprou, D.-H. Kwon, M.~S.
	Mirmoosa, N.~Kantartzis, E.~N. Economou, M.~Kafesaki, C.~M. Soukoulis, and
	S.~A. Tretyakov, ``{Tunable perfect anomalous reflection in metasurfaces with
		capacitive lumped elements},'' in \emph{Proceedings of METAMATERIALS '18},
	2018, pp. 392--394.
	
	\bibitem{Lewi2019}
	T.~Lewi, N.~A. Butakov, and J.~A. Schuller, ``{Thermal tuning capabilities of
		semiconductor metasurface resonators},'' \emph{Nanophotonics}, vol.~8, no.~2,
	pp. 331--338, 2019.
	
	\bibitem{Ju2011}
	L.~Ju, B.~Geng, J.~Horng, C.~Girit, M.~Martin, Z.~Hao, H.~A. Bechtel, X.~Liang,
	A.~Zettl, Y.~R. Shen, and F.~Wang, ``{Graphene plasmonics for tunable
		terahertz metamaterials},'' \emph{Nature Nanotechnology}, vol.~6, no.~10, pp.
	630--634, 2011.
	
	\bibitem{Zhao2015}
	X.~Zhao, K.~Fan, J.~Zhang, H.~R. Seren, G.~D. Metcalfe, M.~Wraback, R.~D.
	Averitt, and X.~Zhang, ``{Optically tunable metamaterial perfect absorber on
		highly flexible substrate},'' \emph{Sensors and Actuators, A: Physical}, vol.
	231, pp. 74--80, 2015.
	
	\bibitem{Yang2016}
	H.~Yang, X.~Cao, F.~Yang, J.~Gao, S.~Xu, M.~Li, X.~Chen, Y.~Zhao, Y.~Zheng, and
	S.~Li, ``{A programmable metasurface with dynamic polarization, scattering
		and focusing control},'' \emph{Scientific Reports}, vol.~6, no. 35692, 2016.
	
	\bibitem{zhao2013tunable}
	J.~Zhao, Q.~Cheng, J.~Chen, M.~Q. Qi, W.~X. Jiang, and T.~J. Cui, ``A tunable
	metamaterial absorber using varactor diodes,'' \emph{New Journal of Physics},
	vol.~15, no.~4, p. 043049, 2013.
	
	\bibitem{Georgiou2018}
	J.~Georgiou, K.~M. Kossifos, M.~A. Antoniades, A.~H. Jaafar, and {N. T. Kemp},
	``{Chua Mem-Components for Adaptive RF Metamaterials},'' in \emph{Proceedings
		of the ISCAS '18}, 2018.
	
	\bibitem{Kan2015}
	T.~Kan, A.~Isozaki, N.~Kanda, N.~Nemoto, K.~Konishi, H.~Takahashi,
	M.~Kuwata-Gonokami, K.~Matsumoto, and I.~Shimoyama, ``{Enantiomeric switching
		of chiral metamaterial for terahertz polarization modulation employing
		vertically deformable MEMS spirals},'' \emph{Nature Communications}, vol.~6,
	pp. 1--7, 2015.
	
	\bibitem{Yu2011a}
	N.~Yu, P.~Genevet, M.~a. Kats, F.~Aieta, J.-P. Tetienne, F.~Capasso, and
	Z.~Gaburro, ``{Light Propagation with Phase Discontinuities: Generalized Laws
		of Reflection and Refraction},'' \emph{Science}, vol. 334, no. October, pp.
	333--337, 2011.
	
	\bibitem{Moccia2017}
	M.~Moccia, S.~Liu, R.~Y. Wu, G.~Castaldi, A.~Andreone, T.~J. Cui, and V.~Galdi,
	``{Coding Metasurfaces for Diffuse Scattering: Scaling Laws, Bounds, and
		Suboptimal Design},'' \emph{Advanced Optical Materials}, vol.~5, no.~19, pp.
	1--11, 2017.
	
	\bibitem{Hosseininejad2019a}
	S.~E. Hosseininejad, K.~Rouhi, M.~Neshat, A.~Cabellos-Aparicio, S.~Abadal, and
	E.~Alarc{\'{o}}n, ``{Digital Metasurface Based on Graphene: An Application to
		Beam Steering in Terahertz Plasmonic Antennas},'' \emph{IEEE Transactions on
		Nanotechnology}, vol.~18, no.~1, pp. 734--746, 2019.
	
	\bibitem{Dong2015}
	D.~S. Dong, J.~Yang, Q.~Cheng, J.~Zhao, L.~H. Gao, S.~J. Ma, S.~Liu, H.~B.
	Chen, Q.~He, W.~W. Liu, Z.~Fang, L.~Zhou, and T.~J. Cui, ``{Terahertz
		Broadband Low-Reflection Metasurface by Controlling Phase Distributions},''
	\emph{Advanced Optical Materials}, vol.~3, no.~10, pp. 1405--1410, 2015.
	
	\bibitem{Zhang2017b}
	L.~Zhang, S.~Liu, L.~Li, and T.~J. Cui, ``{Spin-Controlled Multiple Pencil
		Beams and Vortex Beams with Different Polarizations Generated by
		Pancharatnam-Berry Coding Metasurfaces},'' \emph{ACS Applied Materials and
		Interfaces}, vol.~9, no.~41, pp. 36\,447--36\,455, 2017.
	
	\bibitem{8889025}
	T.~{Shan} and M.~{Li}, ``Coding programmable metasurfaces based on deep
	learning techniques,'' in \emph{2019 IEEE International Symposium on Antennas
		and Propagation and USNC-URSI Radio Science Meeting}, July 2019, pp. 1--2.
	
	\bibitem{8b03171}
	Z.~Liu, D.~Zhu, S.~P. Rodrigues, K.-T. Lee, and W.~Cai, ``Generative model for
	the inverse design of metasurfaces,'' \emph{Nano Letters}, vol.~18, no.~10,
	pp. 6570--6576, 2018, pMID: 30207735.
	
	\bibitem{5033327}
	S.~Inampudi and H.~Mosallaei, ``Neural network based design of metagratings,''
	\emph{Applied Physics Letters}, vol. 112, no.~24, p. 241102, 2018.
	
	\bibitem{Cui2016}
	T.-J. Cui, S.~Liu, and L.-L. Li, ``{Information entropy of coding
		metasurface},'' \emph{Light: Science {\&} Applications}, vol.~5, no.~11, p.
	e16172, 2016.
	
	\bibitem{Liu2016b}
	S.~Liu, T.~J. Cui, L.~Zhang, Q.~Xu, Q.~Wang, X.~Wan, J.~Q. Gu, W.~X. Tang,
	M.~{Qing Qi}, J.~G. Han, W.~L. Zhang, X.~Y. Zhou, and Q.~Cheng,
	``{Convolution Operations on Coding Metasurface to Reach Flexible and
		Continuous Controls of Terahertz Beams},'' \emph{Advanced Science}, vol.~3,
	no.~10, pp. 1--12, 2016.
	
	\bibitem{Liu2018ISCAS}
	F.~Liu, A.~Pitilakis, M.~S. Mirmoosa, O.~Tsilipakos, X.~Wang, A.~C.
	Tasolamprou, S.~Abadal, A.~Cabellos-Aparicio, E.~Alarc{\'{o}}n, C.~Liaskos,
	N.~V. Kantartzis, M.~Kafesaki, E.~N. Economou, C.~M. Soukoulis, and
	S.~Tretyakov, ``{Programmable Metasurfaces: State of the art and
		Prospects},'' in \emph{Proceedings of the ISCAS '18}, 2018.
	
	\bibitem{Liu2017a}
	S.~Liu and T.~J. Cui, ``{Flexible Controls of Terahertz Waves Using Coding and
		Programmable Metasurfaces},'' \emph{IEEE Journal of Selected Topics in
		Quantum Electronics}, vol.~23, no.~4, 2017.
	
	\bibitem{Tasolamprou2018}
	A.~C. Tasolamprou, M.~S. Mirmoosa, O.~Tsilipakos, A.~Pitilakis, F.~Liu,
	S.~Abadal, A.~Cabellos-Aparicio, E.~Alarc{\'{o}}n, C.~Liaskos, N.~V.
	Kantartzis, S.~Tretyakov, M.~Kafesaki, E.~N. Economou, and C.~M. Soukoulis,
	``{Intercell wireless communication in software-defined metasurfaces},'' in
	\emph{Proceedings of the ISCAS '18}, 2018.
	
	\bibitem{Saeed2018b}
	T.~Saeed, C.~Skitsas, D.~Kouzapas, M.~Lestas, V.~Soteriou, A.~Philippou,
	S.~Abadal, C.~Liaskos, L.~Petrou, J.~Georgiou, and A.~Pitsillides, ``{Fault
		Adaptive Routing in Metasurface Network Controllers},'' in \emph{Proceedings
		of the NoCArc '18}, 2018.
	
	\bibitem{Saeed2019}
	T.~Saeed, S.~Abadal, C.~Liaskos, A.~Pitsillides, H.~Taghvaee,
	A.~Cabellos-Aparicio, M.~Lestas, and E.~Alarc{\'{o}}n, ``{Workload
		Characterization of Programmable Metasurfaces},'' in \emph{Proceeding of the
		NANOCOM '19}, 2019.
	
	\bibitem{Pitilakis2018MMParadigm}
	A.~Pitilakis, A.~C. Tasolamprou, C.~Liaskos, F.~Liu, O.~Tsilipakos, X.~Wang,
	M.~S. Mirmoosa, K.~Kossifos, J.~Georgiou, A.~Pitsilides, N.~Kantartzis,
	S.~Ioannidis, E.~N. Economou, M.~Kafesaki, S.~A. Tretyakov, and C.~M.
	Soukoulis, ``Software-defined metasurface paradigm: Concept, challenges,
	prospects,'' in \emph{Proceedings of the METAMATERIALS '18}.\hskip 1em plus
	0.5em minus 0.4em\relax {IEEE}, Aug. 2018.
	
	\bibitem{itu2}
	\BIBentryALTinterwordspacing
	{International Telecommunication Union (ITU)}, ``Report of the cpm to wrc-19,''
	in \emph{World Radiocommunication Conference}, no. March, 2019. [Online].
	Available: \url{https://www.itu.int/md/R15-CPM19.02-R-0001/en}
	\BIBentrySTDinterwordspacing
	
	\bibitem{CST}
	\BIBentryALTinterwordspacing
	``{CST Microwave Studio}.'' [Online]. Available: \url{http://www.cst.com}
	\BIBentrySTDinterwordspacing
	
	\bibitem{BalanisBOOK}
	C.~A. Balanis, \emph{{Antenna Theory: Analysis and Design}}, 3rd~ed., Wiley,
	Ed., 2005.
	
	\bibitem{Liu2018b}
	S.~Liu, T.~{Jun Cui}, A.~Noor, Z.~Tao, H.~{Chi Zhang}, G.~{Dong Bai}, Y.~Yang,
	and X.~{Yang Zhou}, ``{Negative reflection and negative surface wave
		conversion from obliquely incident electromagnetic waves},'' \emph{Light:
		Science and Applications}, vol.~7, no.~5, pp. 18\,008--18\,011, 2018.
	
	\bibitem{Srinivasan2004}
	J.~Srinivasan, S.~Adve, P.~Bose, and J.~Rivers, ``{The impact of technology
		scaling on lifetime reliability},'' in \emph{Proceedings of the DSN '04},
	2004, pp. 177--186.
	
	\bibitem{Zhang2018a}
	L.~Zhang, R.~Y. Wu, G.~D. Bai, H.~T. Wu, Q.~Ma, X.~Q. Chen, and T.~J. Cui,
	``{Transmission-Reflection-Integrated Multifunctional Coding Metasurface for
		Full-Space Controls of Electromagnetic Waves},'' \emph{Advanced Functional
		Materials}, vol.~28, no.~33, pp. 1--9, 2018.
	
	\bibitem{Moccia2018}
	M.~Moccia, C.~Koral, G.~P. Papari, S.~Liu, L.~Zhang, R.~Y. Wu, G.~Castaldi,
	T.~J. Cui, V.~Galdi, and A.~Andreone, ``{Suboptimal Coding Metasurfaces for
		Terahertz Diffuse Scattering},'' \emph{Scientific Reports}, vol.~8, no.~1,
	pp. 2--10, 2018.
	
	\bibitem{Xu2017b}
	H.-X. Xu, S.~Ma, X.~Ling, X.-K. Zhang, S.~Tang, T.~Cai, S.~Sun, Q.~He, and
	L.~Zhou, ``{Deterministic Approach to Achieve Broadband
		Polarization-Independent Diffusive Scatterings Based on Metasurfaces},''
	\emph{ACS Photonics}, vol.~5, pp. 1691--1702, 2017.
	
\end{thebibliography}
\end{document}